# 200,000+ Deep Learning–inferred Periods of Stellar Variability from the All-Sky Automated Survey for Supernovae

Meir E. Schochet[1,2], Penelope Planet[1], Zachary R. Claytor[3], Jamie Tayar[1], and Adina D. Feinstein[2,4]
[1] Department of Astronomy, University of Florida, Gainesville, FL 32611, USA; meir@msu.edu
[2] Department of Physics and Astronomy, Michigan State University, East Lansing, MI 48824, USA
[3] Space Telescope Science Institute, Baltimore, MD 21218, USA

## Abstract

Stars exhibit a range of variability periods that depend on their mass, age, and evolutionary stage. For space-based photometric data, convolutional neural networks (CNNs) have demonstrated success in recovering and measuring periodic variability from photometric missions like Kepler and TESS. All-sky ground-based surveys can have similar if not longer baselines than space-based missions; however, these datasets are more challenging to work with due to irregular sampling, more complex systematics, and larger data gaps. In this work, we demonstrate that CNNs can be used to derive variability periods from ground-based surveys. From the All-Sky Automated Survey for Supernovae, we recover 208,260 variability periods between 1 and 30 days, approximately 60% of which are new detections. We recover periods for active RSCVn, anomalous sub-subgiants, and cool dwarfs that are consistent with previously measured rotation periods, while periods for stars above the Kraft break are generally spurious. We also identify periodic signals in tens of thousands of giant stars that correspond to frequencies of stellar oscillations rather than rotation. Our results highlight that CNNs can be used on sparsely sampled ground-based photometry to recover periodicity. We conclude that the findings of our work are very promising for the potential recovery of hundreds of thousands of stellar rotation periods in data from the Vera C. Rubin Observatory's Legacy Survey of Space and Time and the Nancy Grace Roman Space Telescope's Galactic Bulge Time Domain Survey.

## 1. Introduction

Stellar rotation and intrinsic variability evolve over time. Stars inherit angular momentum from initial cloud collapse, and low-mass stars lose their angular momentum in stellar winds (A. Skumanich 1972; S. P. Matt et al. 2015), similar to our own solar wind (E. N. Parker 1958; E. J. Weber & L. Davis 1967). Comprehensive analyses of single star rotational evolution have demonstrated that many other phenomena drive interior angular momentum transport such as meridional circulation or magnetic instabilities (e.g., A. Maeder & G. Meynet 2000; S. Mathis 2013; C. Aerts et al. 2019). More recent studies have suggested that there is still much to uncover about the rotational evolution of low-mass stars, including truncated braking (J. L. van Saders et al. 2016), interior behavior at the fully convective boundary (F. Chiti et al. 2024), stalled spindown (J. L. Curtis et al. 2020), the effect of spots (L. Cao & M. H. Pinsonneault 2022), core–envelope angular momentum transfer in cool stars (L. Cao et al. 2023), radius inflation (G. Somers & K. G. Stassun 2017), the distribution of rotation rates at birth (C. T. Coker et al. 2016; G. Somers et al. 2017), and the consequences of binarity (J. Tayar et al. 2015; A. Phillips et al. 2023; J. Yu et al. 2024).

Starting with the Sun, observers have used the passage of spots across the line of sight and the resulting change in surface brightness to estimate a solar rotation period (R. Carrington 1863). Dedicated monitoring campaigns can regularly monitor the brightness of stars, allowing observers to understand the rotational properties of stars like the Sun (S. Bhattacharya et al. 2021), young stars (P. Hartigan et al. 2011; A. D. Feinstein et al. 2020; S. T. Douglas et al. 2024), stars in clusters (S. T. Douglas et al. 2017; L. Long et al. 2023; L. Sha et al. 2024), low-mass field M dwarfs (J. Irwin et al. 2011; E. R. Newton et al. 2016), and more-massive stars (J. Sikora et al. 2019), among others. Expansive repositories of photometric light curves from space-based surveys at high cadence and with minimal data gaps, like Convection, Rotation, and planetary Transits (CoRoT; M. Auvergne et al. 2009), Kepler (W. J. Borucki et al. 2010), Kepler Second Light (K2; S. B. Howell et al. 2014), and the Transiting Exoplanet Survey Satellite (TESS; G. R. Ricker et al. 2014) have allowed for the precise measurement of stellar rotation periods. To date, tens of thousands of stars have measured $P_{\rm rot}$ values from space-based photometric monitoring missions including CoRoT, Kepler, K2, and TESS (S. Meibom et al. 2011; L. Affer et al. 2012, 2013; J.-D. do Nascimento et al. 2012; A. McQuillan et al. 2013, 2014; A. R. G. Santos et al. 2019, 2021b; B. L. Canto Martins et al. 2020; T. Reinhold & S. Hekker 2020; E. A. Avallone et al. 2022; R. J. Holcomb et al. 2022; Z. R. Claytor et al. 2024b). However, these catalogs suffer from a limited time baseline determined by the individual telescope monitoring patterns (Z. R. Claytor et al. 2024b) and lifetime (A. R. G. Santos et al. 2019). Future missions—such as the Nancy Grace Roman Space Telescope

[4] NASA NHFP Fellow.

(D. Spergel et al. 2015)—will help expand this existing sample of stars.

Ground-based surveys have been able to measure stellar $P_{\rm rot}$ specifically in stellar clusters (S. Meibom et al. 2011; D. J. Fritzewski et al. 2021; E. M. Cole-Kodikara et al. 2023). However, generalizable studies across the field are challenging due to a combination of difficulties in collecting observations at high enough cadence and over a sufficiently long baseline to measure the range of possible $P_{\rm rot}$. Despite these challenges, surveys such as MEarth (P. Nutzman & D. Charbonneau 2008), the Zwicky Transient Facility (ZTF; E. C. Bellm et al. 2019), and the All-Sky Automated Survey for Supernovae (ASAS-SN; B. J. Shappee et al. 2014; C. S. Kochanek et al. 2017) have fortuitously satisfied both requirements and have measured $P_{\rm rot}$ for ∼100,000 stars (J. Irwin et al. 2011; E. R. Newton et al. 2016; T. Jayasinghe et al. 2019, 2020; M. Pawlak et al. 2019; Y. L. Lu et al. 2022; A. Phillips et al. 2023).

ASAS-SN (B. J. Shappee et al. 2014; C. S. Kochanek et al. 2017) is a ground-based all-sky monitoring survey designed to observe the sky nightly for transient events. ASAS-SN began collecting $V$-band observations of objects with $V_{\rm mag} < 17$ in 2014, and in 2018, all of the units were switched to $g$-band filters that could observe objects with $g_{\rm mag} < 18$ (C. T. Christy et al. 2023). The full ASAS-SN dataset has more than 100 million sources, and the collection includes a substantial number of stellar light curves, some of which possess likely rotation signals that have previously been characterized (T. Jayasinghe et al. 2019, 2020; M. Pawlak et al. 2019).

Traditional measurements of $P_{\rm rot}$ have employed Lomb–Scargle periodograms (N. R. Lomb 1976; J. D. Scargle 1982), autocorrelation functions (A. McQuillan et al. 2013, 2014), wavelet transformations (S. Mathur et al. 2010; R. A. García et al. 2014), and Gaussian processes (R. Angus et al. 2018; T. A. Gordon et al. 2021). However, these methods are computationally intensive and require careful algorithm profiling to ensure efficient computations when applied to catalog sizes of the order of $10^8$ (M. W. Coughlin et al. 2021). On the other hand, machine learning (ML) is a powerful tool that can be used to analyze large data. Deep-learning networks have already demonstrated their utility across a broad range of astrophysical topics such as elemental abundance determination (H. W. Leung & J. Bovy 2019), flare statistics (A. D. Feinstein et al. 2020), galaxy redshift predictions (Q. Lin et al. 2024), and cosmological parameter estimation (J.-Y. Lee et al. 2024) to name a few. While ML models may systematically underestimate extreme values (Y.-S. Ting 2024), these techniques are still a powerful tool that can be used to infer conclusions from large, complicated datasets.

We present a new convolutional neural network (CNN) framework that is capable of measuring stellar variability periods from ASAS-SN light curves, including a significant number of periods that we attribute to rotation. The paper is structured as follows. In Section 2, we describe the ASAS-SN data we use. In Section 3, we discuss our simulated training set and CNN architecture. In Section 4, we present different tests of our CNN to demonstrate the robustness of our method. In Section 5, we validate our inferred $P_{\rm rot}$ by comparing to archival catalogs. In Section 6, we present trends in inferred $P_{\rm rot}$ as a function of different stellar populations. We conclude in Section 7. We provide all of the scripts used to complete this work hosted on GitHub at https://github.com/m-schochet/asas-sn-cnn. We include a GitHub icon (✎) next to figures in this work that link to Jupyter notebook files used to generate that figure, and all notebooks can also be accessed on Zenodo at doi:10.5281/zenodo.17387593.

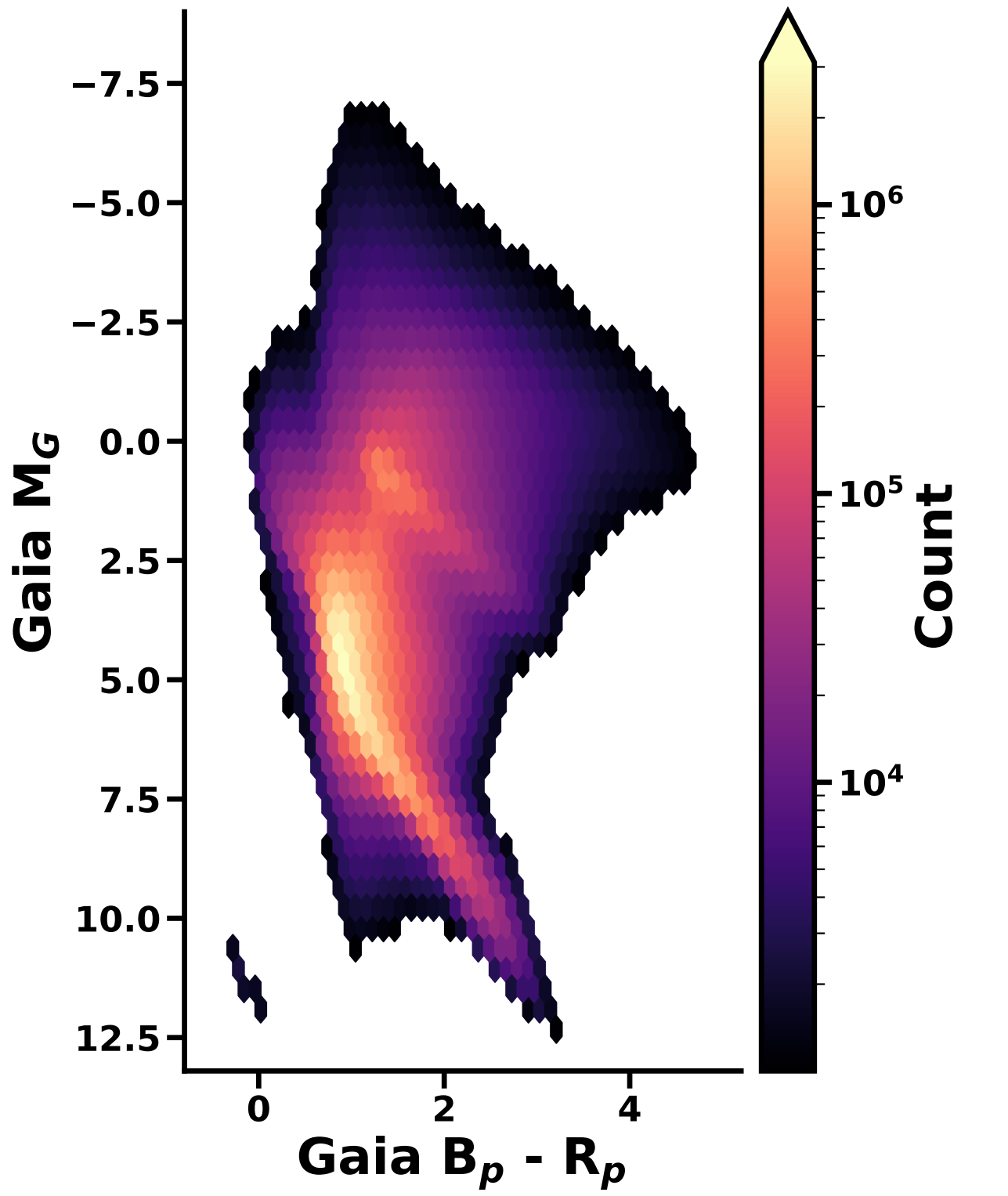

**Figure 1.** A Hertzsprung–Russell (H-R) diagram of stars observed by ASAS-SN and colored using Gaia (Gaia Collaboration et al. 2016, 2018; D. W. Evans et al. 2018). The bins are only filled if they contain 1000 or more stars. Neither our absolute magnitudes nor our photometric colors are extinction corrected. ✎

## 2. ASAS-SN Data

The ASAS-SN survey has a 10 yr baseline of photometric observations for millions of stars (Figure 1). Each ASAS-SN unit has four 14 cm telescopes on a common mount; there are five ASAS-SN units, located in Hawai'i, Texas, South Africa, and two units in Chile. This configuration allows ASAS-SN to observe the entire visible sky every night. We used Globus (I. Foster 2011; B. Allen et al. 2012) to transfer the ASAS-SN catalog from the University of Hawaii to the University of Florida supercomputing servers (a.k.a. HiPerGator).

ASAS-SN data is separated into catalogs based on NASA's High Energy Astrophysics Science Archive Research Center (HEASARC), with the addition of the ASAS-SN **stellar_main** catalog to denote all observed stars. We downselect the stars we analyze using the following criteria:

1. All stars were observed in the same filter. This limits our light curve to observations that were taken after all ASAS-SN units were equipped with $g$-band filters (MJD: 58423).
2. All stars have an identical time baseline. We choose to analyze stars with observations from UT 2019 January 1 (MJD: 58484.5) to UT 2024 January 1 (MJD: 60311.5).
3. The light curves do not have any bad quality observations. We used two metrics to distinguish bad from good photometric points. First, we flagged any data points where the measured magnitude error, $\sigma_{mag} \geqslant 99$, as assigned by the ASAS-SN data reduction pipeline. Second, the ASAS-SN data reduction pipeline labels photometric points as "G" if the observation was "good." We use this flag to remove bad points as well.
4. The light curve possesses $\geqslant$150 photometric observations over the baseline defined in Criteria 2.

We found that ~97% of the full ASAS-SN **stellar_main** catalog passed our four criteria, resulting in a total of 96,304,837 light curves.

## 3. The Convolutional Neural Network

Here we present how we generate our simulated light curves, the transformation applied to our light curves for CNN input, and the architecture of our neural network. From our simulated light curves, we divide the set into our training ("how to set weights"), validation ("how is the model doing, when do we stop training"), and test ("how good is the trained network") sets. The ratio between these sets is 80:10:10 unless noted otherwise.

### 3.1. Data Preparation

The efficiency of a CNN is highly dependent on the training data it is provided, with a potentially greater impact on network performance than even the fine-tuning of hyperparameters (J. Tayar et al. 2023). Due to the observing strategy, cadence, and data quality of ASAS-SN, the data is known to contain systematic noise and strong aliases at 1, 2, and 30 days (T. Jayasinghe et al. 2020). As a result, we need to ensure our training data will incorporate these real ASAS-SN "noise" signals as well as a known $P_{rot}$. To achieve this, we generate a set of synthetic light curves with known $P_{rot}$ and inject them into ASAS-SN light curves that are meant to capture ASAS-SN systematics. These injected light curves constitute our training, validation, and testing sets.

#### 3.1.1. Simulated Rotation Periods

We created a simulated training set following the procedure of Z. R. Claytor et al. (2022, hereafter C22). This provides a good test as to whether simulated training samples—that were previously used to infer $P_{rot}$ from space-based photometry—can also find use with ground-based observations. C22 generated a library of simulated light curves with spot-modulated rotation periods and then "injected" the simulated periodicity into a quiescent sample of sources that possesses the systematic noise of the survey the neural network will predict on.

We simulate light curves with `butterpy` v1.0.0 (Z. Claytor et al. 2024a), which generates and evolves light curves with physically motivated models of stellar spot emergence based on the simplified flux emergence model of S. Aigrain et al. (2015). We simulated 1 million light curves with input parameters sampled from the distributions in Table 1.

#### 3.1.2. ASAS-SN Training Templates

Here we describe our best "quiescent" sample of light curves to inject our `butterpy` models into. C22 injected their `butterpy` simulations into a set of galaxies within the TESS Southern Continuous Viewing Zone to model known TESS systematics, which enabled their prediction of $P_{rot}$ in TESS observations. We tried a similar approach using the ASAS-SN Million Quasar Catalog (E. W. Flesch 2023, hereafter milliquas).

**Table 1**
Distribution of Simulation Input Parameters

| Parameter | Range | Distribution | Symbol |
|---|---|---|---|
| Equatorial rotation period | 1–30 days | Uniform | $P_{eq}$ |
| Activity level | 0.1–10× solar | Log-uniform | A |
| Activity cycle period | 1–40 yr | Log-uniform | $T_{cycle}$ |
| Activity cycle overlap | 0.1 yr–$T_{cycle}$ | Log-uniform | $T_{overlap}$ |
| latitude$_{spot,min}$ | 0°–40° | Uniform | $\lambda_{min}$ |
| latitude$_{spot,max}$ | $\lambda_{min}$ + 5°–85° | Uniform | $\lambda_{max}$ |
| Spot lifetime | 1–10 | Log-uniform | $\tau_{spot}$ |
| Inclination | 0°–90° | Uniform in $\sin^2 i$ | $i$ |
| Magnetic field scaling factor | 0.005 | Fixed | $\alpha_{med}$ |
| Latitudinal rotation shear | 0.1–1 (50%)<br>0 (25%)<br>−1 to −0.1 (25%) | Log-uniform<br>Fixed<br>Log-uniform | $\Delta\Omega/\Omega_{eq}$ |

**Note.** We adopt nearly the same the distributions used by Z. R. Claytor et al. (2022), and use similar values to those introduced in their work. The periods sampled are in a smaller range than expected in nature; however, this range of periods was chosen to provide the best chance of inferring $P_{rot}$ given the irregular cadence and baseline of examination for the ASAS-SN stars.

We used the SkyPatrol Python client `pyasassn` (K. Hart et al. 2023) to download all light curves from the `milliquas` catalog hosted on the SkyPatrol website.[5] We then applied the same quality cuts detailed in Section 2 to each source, resulting in a final template sample of 203,991 milliquas sources. However, we found that this network architecture was incapable of detecting $P_{rot}$ in ASAS-SN stellar data that was not provided to the network during training, indicating that milliquas templates were not an ideal example of ASAS-SN systematics.

Recently, Z. R. Claytor & J. Tayar (2025, hereafter C25) extended their work (C22; Z. R. Claytor et al. 2024b) to the Kepler Bonus dataset (J. Martínez-Palomera et al. 2023). However, due to a lack of quasars and galaxies in the Kepler field of view, the authors used light curves of red clump stars as their templates. These stars are typically slowly rotating (J. Tayar & M. H. Pinsonneault 2018; C. M. Daher et al. 2022; R. A. Patton et al. 2024) and have nondetectable rotation signatures (see, e.g., T. Ceillier et al. 2017). To identify red clump stars, C25 computed the absolute Gaia magnitude for all Kepler stars ($M_G$), where

$$M_G = G - 5\log_{10}(d) + 5\,. \quad (1)$$

In Equation (1), $G$ is the Gaia $G$-band magnitude (`phot_g_mean_mag`), and $d$ is the distance to the object calculated as

$$d\,[\mathrm{pc}] = \left|\frac{1000}{p}\right|. \quad (2)$$

In Equation (2), $p$ is the parallax given in milliarcseconds. Using these absolute Gaia magnitudes, red clump stars were identified as stars with $-0.5 < M_G < 1.5$.

[5] http://asas-sn.ifa.hawaii.edu/skypatrol/

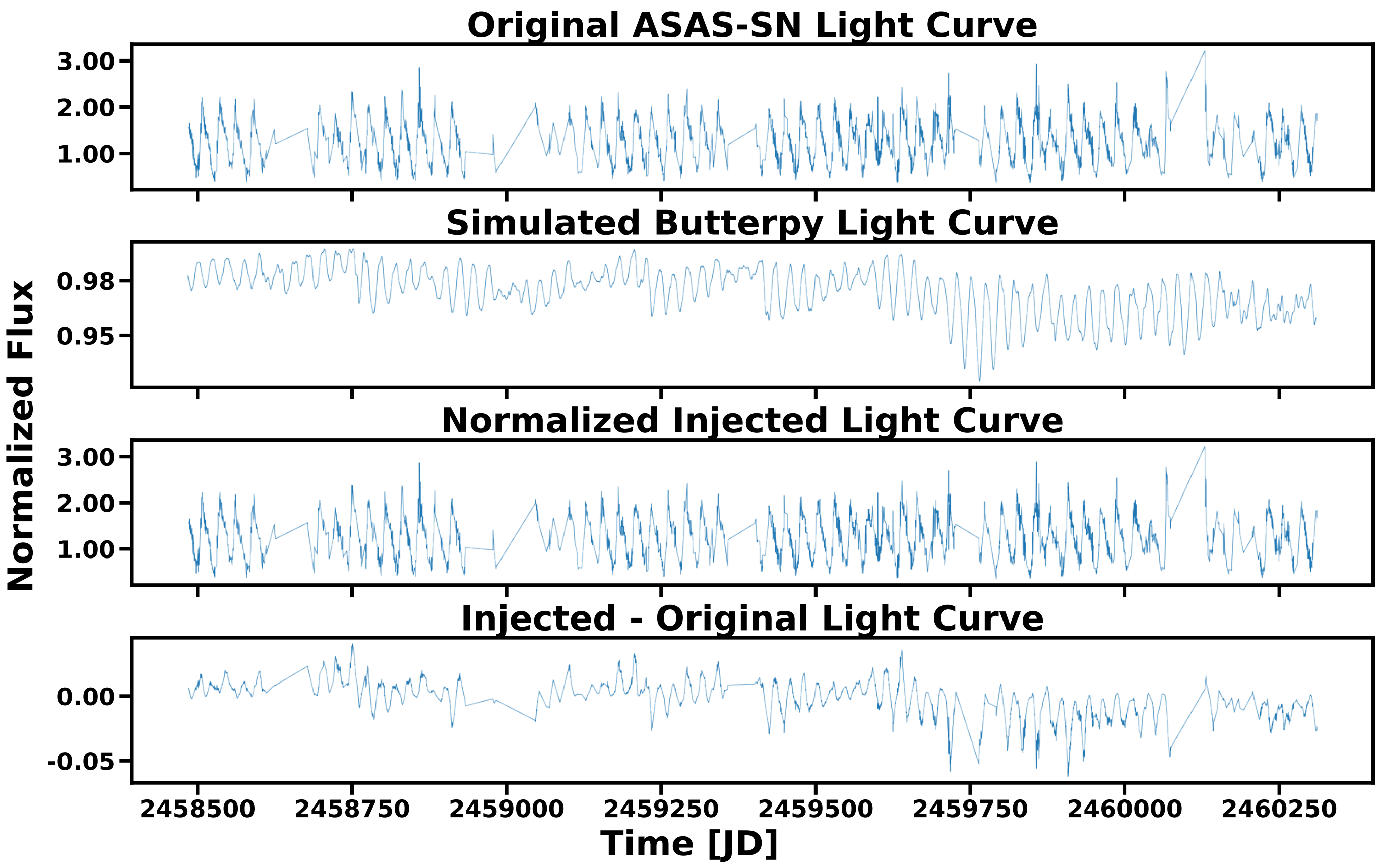

**Figure 2.** The light curve of ASASSN-V J050623.37-712251.8. Gaps in the data are intrinsic to ASAS-SN light curves. First row: original light curve. Second row: an example `butterpy` simulation over the same baseline. The complex sinusoidal pattern demonstrates the intricate physics that `butterpy` simulations can generate. Third row: a `butterpy` model injected into a noise template. Fourth row: the "injected" light curve minus the original noise template to demonstrate that we can recover the injected periodicity and the expected observing gaps.

We developed a secondary training set using red clump stars as our ASAS-SN templates. We found there are of the order of 10,000,000 sources in the **stellar_main** catalog with $-0.5 < M_G < 1.5$, which is a sufficient number of templates to curate a training set of an appropriate size for our dataset. Ultimately, we used 1,000,000 red clump stars as the ASAS-SN templates into which we injected `butterpy` periodic signals, mostly due to computational limitations. Where there are large uncertainties in the parallax, this estimate of $M_G$ may be offset from the true absolute magnitude. In particular, we note that we have not applied the zero-point correction from X. Luri et al. (2018) and have not accounted for the Bayesian uncertainty that is necessarily included in any Gaia astrometric measurements. However, even in cases of large astrometric uncertainty (of the order of 0.5 mas and −0.029 mas for the zero-point offset), we would expect a difference in measured versus absolute magnitude of the order of ∼3 for the most significantly affected sources. Considering that we are targeting red clump giants that are bright and nearby (C. Soubiran et al. 2003), we do not expect significant issues with the astrometric uncertainty. These errors should not affect our selection of appropriate templates, as any stellar light curve will necessarily incorporate the systematic noise from ASAS-SN that we are interested in teaching our CNN to ignore. Thus, we use a combination of milliquas and red clump star templates when generating our training, validation, and testing sets, and we explore the specifics further in Section 4.2.

We note that when originally curating the red clump sample, there were 1105 repeat templates. This propagated to 2210 objects in our training set having a nonunique red clump template. However, the repetition of training templates does not significantly influence the predictive efficiency of our CNN (Section 4.4). Even so, we note that three stars that we report $P_{\rm rot}$ for were contained in our training set and repeated once. Those targets are TIC 388043164 ($P_{\rm rot,pred} = 18.805 \pm 2.45$ days), TIC 397920546 ($P_{\rm rot,pred} = 24.962 \pm 6.23$ days), and TIC 461382691 ($P_{\rm rot,pred} = 25.364 \pm 4.11$ days). Overall, the slight repetition of a few templates does not impact the results of the CNN.

### 3.2. Injection

We combine our synthetic `butterpy` light curves and our templates using the following algorithm. First, we linearly interpolate the `butterpy` light curve to the time steps of the ASAS-SN template light curve. Next, we piecewise multiply each ASAS-SN flux value to the interpolated `butterpy` flux at those time steps. Finally, we normalize the resulting light curve by the median value of `butterpy` fluxes. We describe the full algorithm in Appendix A. We chose the median value, as it is more robust against outliers. We applied this algorithm across our entire template sample (milliquas and red clump stars). This gave us our "injected" flux training set. This final set contains periodicity from the simulated `butterpy` signal and systematics from ASAS-SN. We show an example of this in Figure 2.

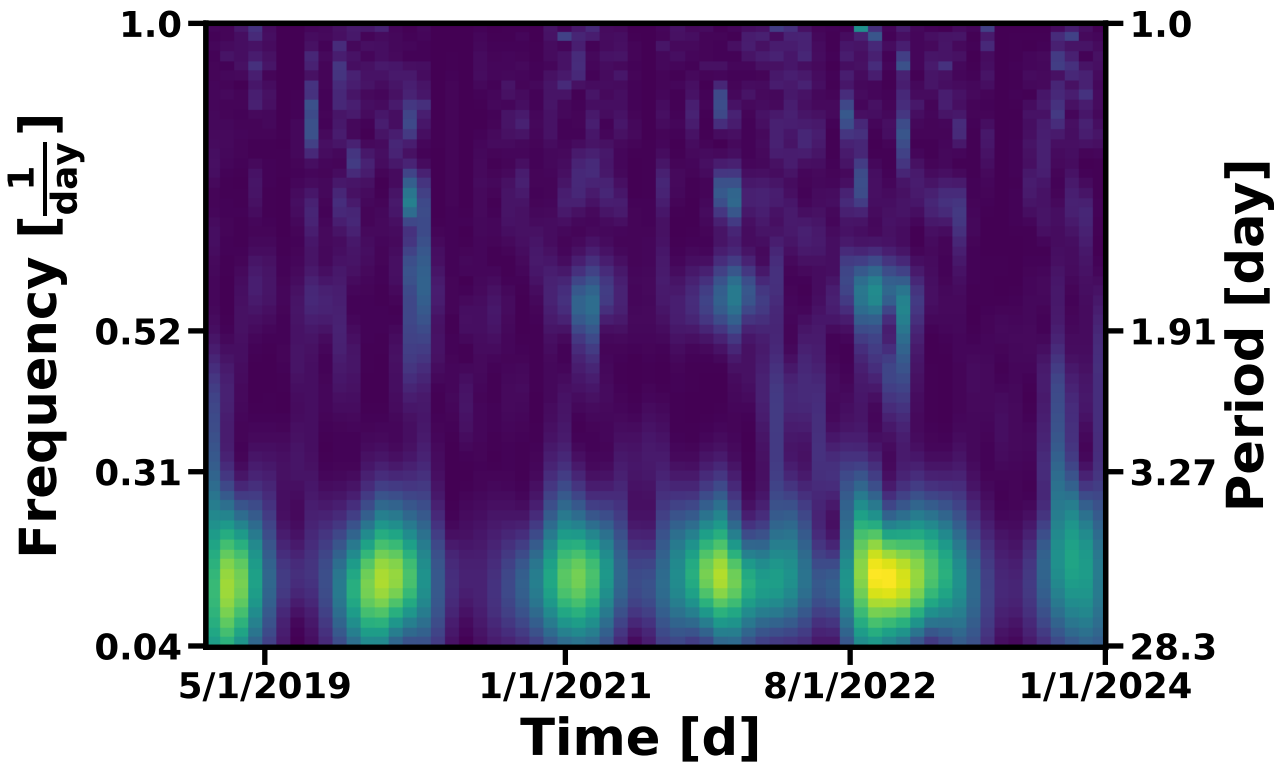

**Figure 3.** An example signal transformation applied to ASASSN-V J050623.37-712251.8, a known spotted rotational variable (T. Jayasinghe et al. 2018). We interpolate over empty regions of the transform using a cubic interpolation. The vertical striping is indicative of noise. Our neural network recovers a period of 27.3 ± 2.23 days, which is consistent with the 26.82 days period from T. Jayasinghe et al. (2018).

### *3.3. Signal Transformation*

Following the injection of the `butterpy` simulated signals, we apply a signal transformation to each light curve to convert them into 2D images that display frequency information across the temporal domain. CNNs learn features from their respective training datasets, with effective predictions in the astrophysical regime demonstrated by networks working with 1D data (M. Hon et al. 2017; W. Liu et al. 2019), or more commonly on 2D "images" (W. W. Zhu et al. 2014; B. Hoyle 2016; A. K. Aniyan & K. Thorat 2017; E. J. Kim & R. J. Brunner 2017; J. Wilde et al. 2022). While studies have shown that periodicity can be learned from 1D stellar light curves (S. Iglesias Álvarez et al. 2023), these results have relied on the regular cadence of missions like Kepler. As a result, it is uncertain whether the irregular cadence and noisy ground-based systematics of ASAS-SN are ideal for this type of architecture.

Following the procedure of C22 for handling gaps and systematics in TESS, we apply a signal transformation to convert our light curves from time series into 2D images that display frequency information as a function of time. This is done through the application of a Lomb–Scargle periodogram (Press & Rybicki algorithm, W. H. Press & G. B. Rybicki 1989) across sections of the light curve and across a range of frequencies that outputs a 2D image that approximates a modified wavelet power spectrum (C. Torrence & G. P. Compo 1998).

To perform this transformation, we use the `pyasassn.LS_wavelet` function.[6] This function takes time series data and associated uncertainties, and runs through two nested for-loops, first in frequency and then in time space. In each frequency loop, a "time step" ($\Delta t$) is created that is proportional to the frequency being evaluated to maintain the "wavelet" structure at all frequencies. Then, inside the time loop, the function generates a Gaussian-modulated sinusoid window that is shifted in time (to overlap on the section of the light curve being evaluated) and divided by the $\Delta t$ created earlier to maintain the wavelet structure. The window is then divided out of the uncertainties—which, in principle, results in the uncertainties nearest to the time step being evaluated to be up-weighted—after which all noninteger uncertainties are masked. The algorithm then performs a Lomb–Scargle periodogram (N. R. Lomb 1976; J. D. Scargle 1982) across the entire light curve, excluding masked points. The light curve's power at the frequency being evaluated is then measured and scaled to the integral of our window function integrated over the $\Delta t$ that we created $\left(\frac{\text{power}}{\sqrt{2\pi}\,\Delta t}\right)$. This formulation leads us to refer to this algorithm as the "chunky LSP," although the output of this function is a 2D array that is analogous to a wavelet transformation (see Appendix B for a programmatic representation of this algorithm). We generate a time grid of length 128 with evenly spaced intervals over our baseline of analysis (MJD 58484.5-60311.5) along with a frequency grid of length 128 using evenly spaced periods ranging from 1–30 days. Any star with a period under 30 days will have undergone more than 60 rotations over our baseline.

Some light curves have intervals of time with fewer than one measurement per day. When performing the chunky LSP on these sections of the light curve, the returned power at low frequencies can be a number of erroneous values including $\pm\infty$ or `NaN`. This is a problem for feeding the transformations into the neural network because we require a common scale of values to ensure uniformity in the data the CNN sees. The simplest common scale for 2D images is to normalize each pixel to an 8 bit integer in the range [0, 255], but this is impossible if a value of $\pm\infty$ or `NaN` appear in the transformation. To circumvent this, after performing the chunky LSP, we perform 2D cubic interpolation with `scipy.interpolate.griddata` to fill in these values. An example of what this transformation looks like is shown in Figure 3. Additionally, the 2D grid interpolation falls back to filling grid points with an `NaN` value if there are no real values within the cubic region with which to interpolate. In the cases where this does occur, we additionally mask over these `NaN` values and replace them with 0; however, we explore the impact of this choice further in Section 4.5.

### *3.4. CNN Architecture*

Our CNN uses a similar architecture to that presented in C22, and used in Z. R. Claytor et al. (2024b, hereafter C24) to predict periods from noisy space-based data. We build our network using the PyTorch python package (`torch`; A. Paszke et al. 2019) using the network architecture in Table 2. Our network uses a series of convolutional layers with rectified linear unit (ReLU) activation followed by time dimension max-pooling. ReLU is a nonlinear activation function with the form

$$f(x) = \max(0, x), \tag{3}$$

which allows for quick learning, as it outputs the input if positive or 0 if negative.

We chose kernels in the convolutional and pooling layers to ensure equivariance in the frequency domain and translational invariance in the time domain. This means that we prevent pooling in the frequency dimension—as that is what we are predicting—and we ensure that the periodicity in the transformation can be identified regardless of where in the time dimension of the transformation it is found (for more details, see I. Goodfellow et al. 2016). The output of the final convolutional layer is then flattened and passed through three

[6] The `pyasassn` Python client can be accessed at https://github.com/asas-sn/skypatrol.

**Table 2**
Convolutional Neural Network Architecture

| Layer Type | Number of Filters[a] | Filter Size | Stride | Activation | Dropout | Output Size |
|---|---|---|---|---|---|---|
| Input image | ⋯ | ⋯ | ⋯ | ⋯ | ⋯ | 64 × 64 |
| Conv2D | 8 | 3 × 3 | 1 × 1 | ReLU | ⋯ | 62 × 62 × 8 |
| MaxPool2D | 1 | 1 × 3 | 1 × 3 | ⋯ | 10% | 62 × 20 × 8 |
| Conv2D | 16 | 3 × 3 | 1 × 1 | ReLU | ⋯ | 60 × 18 × 16 |
| MaxPool2D | 1 | 1 × 3 | 1 × 3 | ⋯ | 10% | 60 × 6 × 16 |
| Conv2D | 32 | 3 × 3 | 1 × 1 | ReLU | ⋯ | 58 × 4 × 32 |
| MaxPool2D | 1 | 1 × 4 | 1 × 4 | ⋯ | 10% | 58 × 1 × 32 |
| Flatten | ⋯ | ⋯ | ⋯ | ⋯ | ⋯ | 1856 |
| Dense | ⋯ | ⋯ | ⋯ | ReLU | 10% | 256 |
| Dense | ⋯ | ⋯ | ⋯ | ReLU | 10% | 64 |
| Output (Dense) | ⋯ | ⋯ | ⋯ | Softplus | ⋯ | 2 |

**Note.** We model the structure of C22 and use three 2D convolutional layers with ReLU activation, max-pooling, and 10% dropout. The output goes through a series of fully connected layers—also using ReLU activation and 10% dropout—and with softplus output. We also use the Adam optimizer (D. P. Kingma & J. Ba 2014) with negative log-Laplacian loss to predict rotation periods with uncertainties.

[a] We explore how varying the number of convolutional filters affects predictive efficiency by testing four different filter architectures when training our network (see Section 4.1). This column displays the number of filters for the network whose inferred periods are reported in this work.

fully connected layers, also with ReLU activation. We use a dropout of 10% in the max-pooling and fully connected layers, which randomly assigns a percentage of neurons to 0 in training, thus removing their contributions from the network. This ensures that learning is focused on more generalized features. The final fully connected layer used Softplus activation, which has the form

$$f(x) = \ln(1 + e^{x}). \tag{4}$$

Softplus is a smooth approximation of the ReLU function, which ensures that this final layer preserves differentiability as well as requiring a positive output.

We use the Adam optimizer (D. P. Kingma & J. Ba 2014), which allowed our network to vary the learning rate during training, and used a negative log-Laplacian likelihood for our loss function. This loss function allows for the output of both a prediction ($P_{\rm pred}$) and an error ($\sigma$)

$$\mathcal{L} = \ln(2\sigma) + \frac{|P_{\rm true} - P_{\rm pred}|}{\sigma}. \tag{5}$$

We caution against interpreting $\sigma$ as a statistically rigorous estimate of the error on our predictions, but rather as a rough estimate of the accuracy. While in principle $\sigma$ can be used to select a subsample of predictions that can be considered "reliable" (such as in C25), we eventually decided to use the metric of fractional uncertainty for this work to report our "good" sample. Fractional uncertainty in our predictions is calculated as

$$\text{Fractional Uncertainty} = \frac{\sigma}{P_{\rm pred}}, \tag{6}$$

and we discuss in more depth the decision to select our reported sample of periods using this metric over a $\sigma$ cut in Section 5.1.

### *3.5. CNN Training*

Our full partitioned dataset was imported as a single object, and during training, we fed the transformations into the network in batches. We used batch sizes of 100, 500 possible total training epochs, 20 epochs for "early stopping patience," and a learning rate of $10^{-5}$. Varying the batch size only changes the relative speed of the CNN training. Our learning rate was derived from C22. We explore the value of our "early stopping patience" further in Section 4.1.

The training set was 80% of the full dataset, and it was used to determine the model's weights. The validation set, which consisted of 10% of the dataset, was used to determine the early stopping of training. In practice, this meant that we ceased training the network when the average loss on the validation set did not improve over a window of training epochs determined by the "early stopping patience" hyperparameter. At this point, the model is considered trained, and it is tested on the testing set made up of the remaining 10% of the data. A demonstration of the network effectively minimizing loss over the course of training is shown in the top of Figure 4.

While training, the network reports the period predicted for each object. We can compare that to the assigned period from the `butterpy` simulations to determine their reliability, defined as

$$\frac{|P_{\rm pred} - P_{\rm true}|}{P_{\rm true}}. \tag{7}$$

Figure 5 shows the results for both the full sample (top), and for those with fractional uncertainties $\leqslant$25%. We note here that the lack of predictions at very short periods is not novel, and this trend has been shown in identical networks applied to TESS and Kepler Bonus data (C22; C24; C25). Our network only reliably recovers periods for a small fraction of light curves from the whole test set (only 26% have reliability $<$10%). For periods with fractional uncertainty $<$25%, 71% of periods are reliable to within 10%, and 88% are reliable to within 20%; we discuss the distribution and reliability of recovered periods more in Section 5.

## 4. Varying the CNN Inputs

The outputs of any machine learning algorithm are highly sensitive to the data it was trained on. Beyond this, we wanted to test whether an identical architecture used on space-based photometry (C22; C25) can be used on noisier ground-based photometry. To further validate the results of our presented architecture, we perform several tests to examine how varying our simulated light curves and hyperparameters affects the recovery rate of $P_{\rm rot}$.

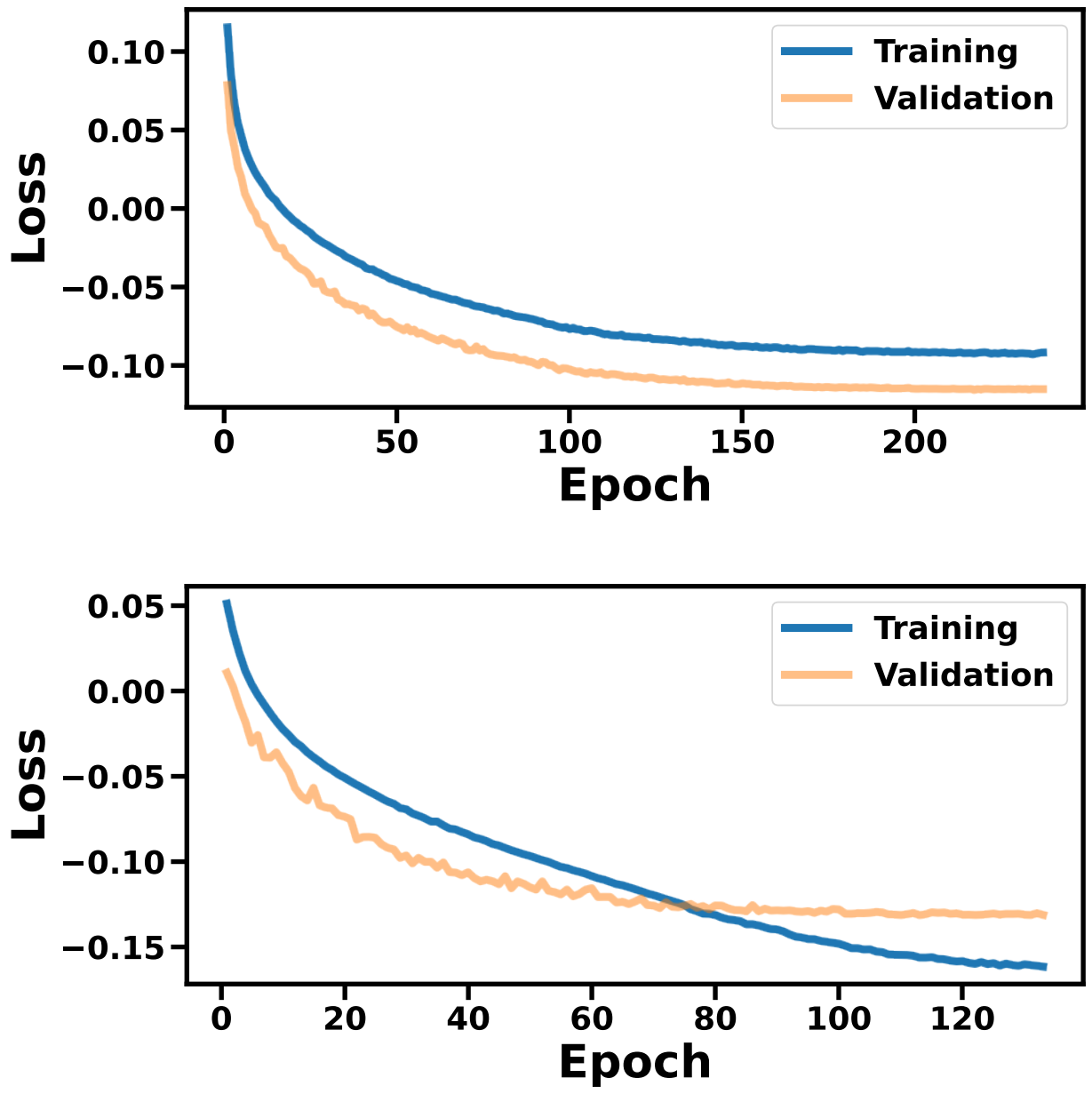

**Figure 4.** The network's loss output for both training and validation sets. Once the network cannot reduce loss output over a number of epochs (set by the early stopping patience parameter), training is ceased. We demonstrate this trend for both a well-fit network (top, row 5 in Table 3) as well as a network that overfit the training data (bottom, row 8 in Table 3). Overfitting can be identified if the training loss continues to decrease while the validation loss had many epochs previously began to asymptote.

Through various modifications of either our training set or network architecture, we aim to address the following questions:

1. How does the number of convolutional filters affect our network's output periods? (Section 4.1; Rows 1-4 in Table 3; Figures 6(a)–(d).)
2. Are stars or quasars/galaxies the best noise template for injection to teach our network to infer rotation periods? (Section 4.2; Rows 1–8 in Table 3; Figures 6(a)–(h).)
3. How many training templates is the ideal number for teaching our network? (Section 4.3; Rows 1 and 9 in Table 3; Figures 6(a) and (i).)
4. Does repeating noise templates in the training set substantially worsen our predictions? (Section 4.4; Rows 9–10 in Table 3; Figures 6(i) and (j).)
5. Does masking over areas of our transformation with minimum non-`NaN` values improve our predictions over masking with `NaN`s? (Section 4.5; Rows 1 and 11 in Table 3; Figures 6(a) and (k).)
6. Does scaling the signal of our transformation improve our predictions? (Section 4.6; Rows 1 and 12 in Table 3; Figures 6(a) and (l).)

To address all of these questions, we use the milliquas light-curve templates with injected `butterpy` $P_{\rm rot}$. This minor difference in training sets does not affect the qualitative conclusion from each test performed. All other properties of the CNN remained the same as described in Section 3.5, unless otherwise noted. We note that we use the fractional uncertainty (Equation (6)) as our metric for evaluating the results of each test. We report the percentage of $P_{\rm rot}$ in the test set that the network reported as accurate to within a fractional uncertainty of 25%. We provide an overview of our changes for each test in Table 3, and a visual summary of the results from each test in Figure 6.

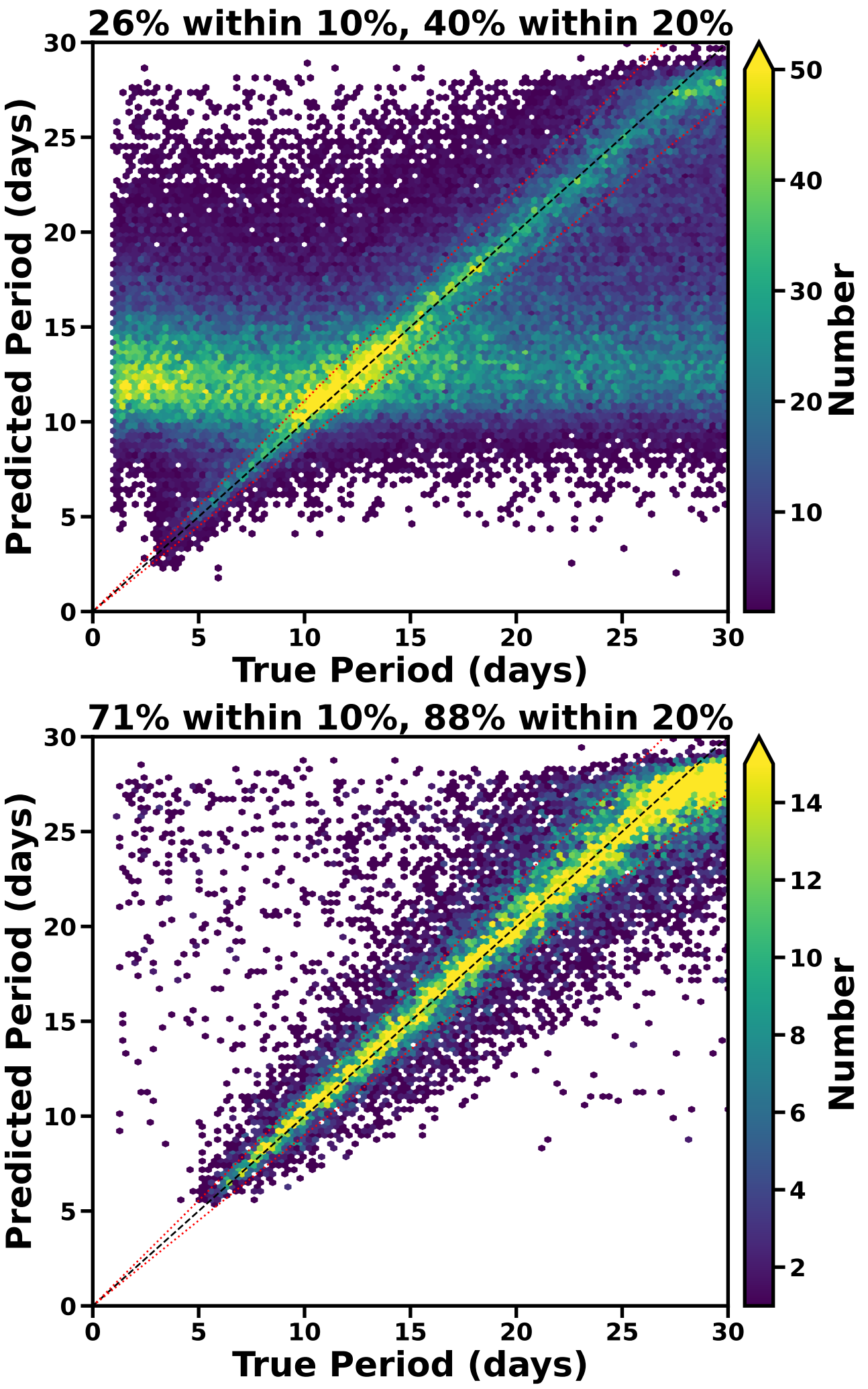

**Figure 5.** The top panel compares the predicted and true periods for the entire test set. The dotted black line is a 1:1 trend of $P_{\rm pred}$: $P_{\rm true}$, while the red dotted lines correspond to 10% differences (as in Equation (7)). The horizontal band at predicted periods of 10–15 days represents a bias toward the probed distribution median, here at ∼15 days (see Figure 5 in C22 for a similar trend). The bottom panel is identical to the top panel, but only includes predictions with fractional uncertainties (Equation (6)) $\leqslant$25%. This shows that both (i) the predicted fractional uncertainty is a good indicator of a successfully predicted period, and (ii) the inferred periods are highly accurate when the fractional uncertainty is <25%. Note that the color bar scaling between the panels is different.

We recognize that there are different architecture and training set configurations that can be applied to address all of the aforementioned questions. Our aim is not to provide a comprehensive overview of these different configurations, but rather to validate the decisions made in our CNN presented in Section 3.

### 4.1. Convolutional Filter Sizes

We want to identify the ideal number of convolutional filters to use when predicting $P_{\rm rot}$. C22 did not identify any major differences between predictions made by CNNs where the

**Table 3**
Convolutional Neural Network Tests

| Figure 6 Panel | Section | Training Set Size | Template Catalog | Repeated Templates | Scaling | Masking Method | Number of Convolutional Filters | Predictive Efficiency |
|---|---|---|---|---|---|---|---|---|
| (a) | 4.1, 4.2, 4.3, 4.5, and 4.6, | $10^6$ | milliquas | Yes | Normal | Nanmask | 8/16/32 | **3.81%** |
| (b) | 4.1 | $10^6$ | milliquas | Yes | Normal | Nanmask | 16/32/64 | **4.35%** |
| (c) | 4.1 | $10^6$ | milliquas | Yes | Normal | Nanmask | 32/64/128 | **4.63%** |
| (d) | 4.1 | $10^6$ | milliquas | Yes | Normal | Nanmask | 64/128/256 | **4.56%** |
| (e) | 4.2 | $10^6$ | clump | No (∼99%[a]) | Normal | Nanmask | 8/16/32 | **19.82%** |
| (f) | 4.2 | $10^6$ | clump | No (∼99%[a]) | Normal | Nanmask | 16/32/64 | **21.54%** |
| (g) | 4.2 | $10^6$ | clump | No (∼99%[a]) | Normal | Nanmask | 32/64/128 | **21.43%** |
| (h) | 4.2 | $10^6$ | clump | No (∼99%[a]) | Normal | Nanmask | 64/128/256 | **21.79%** |
| (i) | 4.3 and 4.4 | 203,991 | milliquas | Yes | Normal | Nanmask | 8/16/32 | **0.18%** |
| (j) | 4.4 | 203,991 | milliquas | No | Normal | Nanmask | 8/16/32 | **0.97%** |
| (k) | 4.5 | $10^6$ | milliquas | Yes | Normal | Minmask | 8/16/32 | **0.19%** |
| (l) | 4.6 | $10^6$ | milliquas | Yes | 10× | Nanmask | 8/16/32 | **2.83%** |

**Note.** This table demonstrates the different predictive efficiencies of the networks tested in Section 4. Our tests demonstrate that the reductions of the training set from $10^6$ to $\sim10^5$ reduced predictive efficiency several times over. Meanwhile, the majority of our tests revealed only marginal improvements while demonstrating other prominent complications over the network whose parameters are shown in the first row of this table. The first column refers to the location of the scatterplot demonstrating this network's predictive efficiency in the summary Figure 6. The bold formatting emphasizes that predictive efficiency is the main indicator of whether a test improved or worsened our network's performance.
[a] See Section 3.1.2 for more details.

number of convolutional filters in the first `Conv2D` layer varied from 8–64. We assess whether or not this is true with ground-based photometry as well. We trained four unique CNNs with varying filter sizes of 8 (16, 32, 64), 16 (32, 64, 128), and 32 (64, 128, 256) in the first, second, and last convolutional layers, respectively.

We trained and predicted on the same examples for all four networks. The results are presented in Figure 6(a), (b), (c), and (d). While the overall predictive efficiency between these models differs by ∼1%, we find that our networks trained on the largest number of convolutional filters were being overfit (bottom of Figure 4) compared to our smaller networks (top of Figure 4), despite using identical hyperparameters. This means that our larger networks continue adjusting weights long after the output predictions stopped “improving.” This could be because the greater number of filters encourages the CNN to identify more complicated trends that our data does not possess. As a result, this test validates our initial choice of convolutional filter sizes from Table 2. While each architecture produces a high-fidelity catalog, we choose to report the catalog from the CNN with the fewest convolutional filters.

### 4.2. Choice of Quiescent Source

We explore the quiescent templates used in our training set. For all of the tests presented in Section 4, we use the milliquas data as our noise templates. However, in our best-trained CNN, our noise templates were instead red clump stars (Section 3.1.2). Here, we aim to assess which “quiescent” source provides a better template for capturing ASAS-SN noise properties and systematics.

We use the same red clump selection protocol defined in Section 3.1.2 (C25). We randomly chose $10^6$ sources, without repetition, from our clump sample to train a new CNN and compare to the milliquas-only CNN, which has repetitive templates. We also repeat this exercise for different numbers of convolutional layers. We had each CNN predict on their test sets. The results of this test are shown in Figures 6(a)–(h).

Our results demonstrate that the CNN's ability to infer reliable $P_{\rm rot}$ in its testing set improved more by than 500% by changing the training set from milliquas to red clump stars. For the purpose of predicting rotation periods from ASAS-SN stars, this test validates our decision to use clump templates as opposed to milliquas. There are several possible hypotheses for why this may be, including the fact that quasars are typically faint and intrinsically variable sources. In addition, when the ASAS-SN pipeline performs aperture photometry on a source, a 2 pixel annulus (∼16″; see K. Hart et al. 2023) is used. For stellar sources, this should not dramatically affect our measurements from night to night, since they are point sources; however, milliquas sources are basically all extended. Again, the results of this work differs from those presented in C22, where training on extended source templates was sufficient; although, this may be due to the larger pixel scale present on TESS (∼21″ as opposed to ∼8″ in ASAS-SN). Our general conclusion is that CNNs should be trained on template data that is as close to the data to be predicted on as possible.

#### 4.2.1. Alternative Templates

In developing our CNN, we relied on using a combination of milliquas and red clump light curves for our training, validation, and testing sets. While this worked for our network, we note that there are other templates one could use. To start, we considered using stars in ASAS-SN with previously measured $P_{\rm rot}$ (T. Jayasinghe et al. 2019, 2020; M. Pawlak et al. 2019). However, these $P_{\rm rot}$ were recovered using a different technique (a random forest classifier followed by a Lomb–Scargle periodogram; see T. Jayasinghe et al. 2018, 2019). As a result, should we train our CNN on these catalogs, it is possible the CNN would imprecisely learn, if the Lomb–Scargle periods were inconsistent from the canonical $P_{\rm rot}$. Alternatively, we considered using ASAS-SN stars with measured $P_{\rm rot}$ from other missions that could be crossmatched (e.g., MEarth, ZTF, or TESS). However, an incomplete set of training, validation, and testing data could cause our CNN to preferentially detect a subset of $P_{\rm rot}$. This would artificially limit our ability to detect new $P_{\rm rot}$. Furthermore, some of the $P_{\rm rot}$ that we would utilize for this method were also derived from machine learning techniques (C24), which may be incomplete or include spurious $P_{\rm rot}$. Determining the ideal template set for our CNN architecture should be investigated, but is beyond the scope of this work.

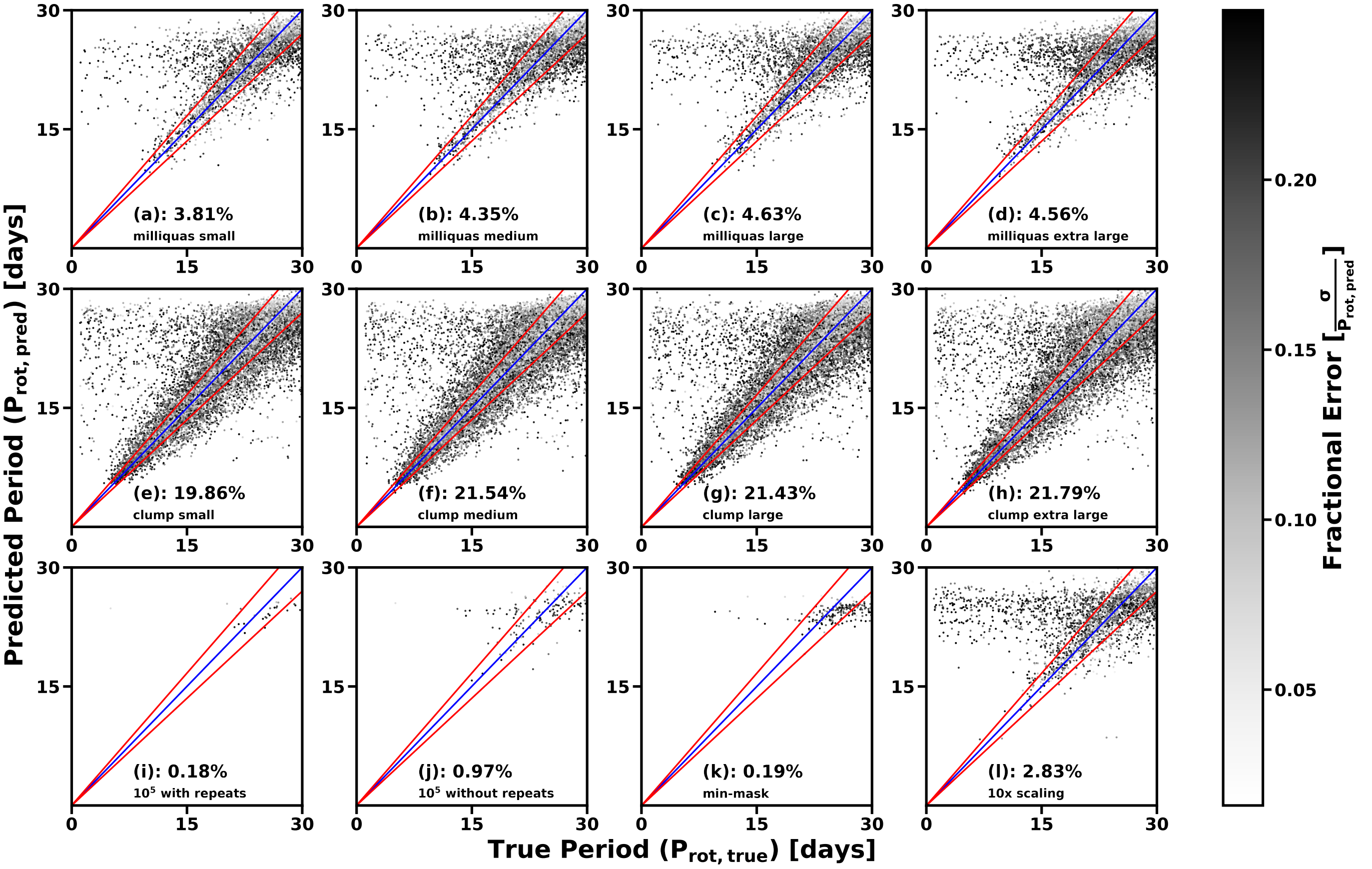

**Figure 6.** The predictive efficiencies of different neural network architectures and training sets. The subpanels correspond to the rows in Table 3. Figures are labeled by "small," "medium," "large," and "extra large" to denote the network "size" when we vary the number of filters. We find there is a negligible difference in inferred $P_{\rm rot}$ when we change the filter sizes. However, we highlight the significant improvement (500% increase) in the predictive power of our network when training on red clump star templates compared to the milliquas templates, which are shown in the first row. The low recovery of periods in (i) and (j) demonstrates that a network trained on $10^5$ templates is ineffective at producing confident predictions when compared to a network trained on $10^6$ templates. The minor improvement in (j) is due to using nonrepeated templates. Similarly, the low recovery in (k) and (l) demonstrates that neither the "min-mask" method of interpolation or the 10× scaling of the signal improved the quantity or quality of predictions compared to our base procedure.

### 4.3. Training-set Size

We explore how many training templates are required to sufficiently distinguish between true $P_{\rm rot}$ and noise. C22 used a training set of size $10^6$ examples. However, other astrophysical CNNs have demonstrated effective training on sets $\sim 10^5$ examples (J. Bialopetravičius et al. 2019; C. J. Burke et al. 2019; N. Monsalves et al. 2024). To explore how well our CNN performs when trained on a smaller sample, we repeat our training using set sizes of $10^5$ and $10^6$.

Our 1 million `butterpy` simulations were then matched up to a milliquas template one-to-one, and segmented into training, validation, and testing sets with the typical 80:10:10 ratio. Additionally, we ensured that repeated milliquas light curves only appeared in either the training, validation, or testing sets. We downselected $10^5$ examples to create the smaller training set; this subset had examples with repetitive milliquas templates. For the smaller set, we divide the sets using a ratio of ∼82:10:8; the reason for this is that we made a calculation error when creating our milliquas training set that caused our testing set to be slightly smaller and our training sample to be slightly larger. However, the $\sim 10^4$ light curves that remain in the validation and testing sets should still have enough variance to allow the network to sample the entire parameter space.

After training networks on both sets, we predicted on their associated test sets to determine their accuracy. These results are presented in Figures 6(a), (i). We find that the predictive efficiency increases from 0.18% to 3.81% when increasing the training set from $10^5$ to $10^6$. This test helped us determine that there is a clear improvement on predictions from networks trained on $10^6$ light curves, despite template repetition used in generating the training set. This validates our use of $10^6$ examples in our training dataset for our CNN presented in Section 3.

### 4.4. Repetition of Templates

Due to the limitations of our templates, we had to repeat examples to sufficiently test how sample size affects the CNN predictive efficiency. However, the resulting predictive efficiency could have improved either due to the increased training set size or if it was able to learn better with repeated templates. Here, we develop a test to distinguish between these possibilities.

To explore the difference in predictive efficiency of a set of $10^5$ examples, we train on a set of $10^5$ examples with and without template repetition. We then predicted on their associated test sets, and the results are presented in Figures 6(i), (j). We find that the predictive efficiency increases from 0.18% to 0.97% when training on examples that do *not* repeat milliquas templates. This improvement, albeit small, supports the use of unique training templates when possible.

### 4.5. Masking Procedure

We examined the ideal procedure for the cubic interpolative masking (Section 3.3) and aim to assess how our masking routine impacts the predictive efficiency of our CNN. In our initial procedure, we masked over all interpolated values of `NaN` or $\pm\infty$ to 0. However, this change may wash out low signal areas of the transformation, resulting in an artificial handicap on the periods we could detect. Here, we test a secondary procedure in which we replaced values of `NaN` and $\pm\infty$ with the minimum non-`NaN` value from the transformation. We trained two CNNs, with the first using our initial reassignment prescription and the second network using the aforementioned "min-mask" method.

We predicted on their associated test sets. The results are presented in Figures 6(a) and (k). We find that the predictive efficiency decreases from 3.81% to 0.19% when implementing the "min-mask" method. This test demonstrates that our CNN is able to better distinguish between $P_{\rm rot}$ and noise when using our initial methodology of replacing NaN/$\pm\infty$ values with 0 in the transformation.

### 4.6. Scaling of Periodic Signal

We choose to explore whether the low predictive capabilities of our previously tested CNNs were due to a lack of relevant signal to identify. We examine the effect of scaling of our transformations prior to training. The idea of scaling would be to amplify the low-power sections of the transformation, which may include true $P_{\rm rot}$ information. By scaling the background to remove low-amplitude signals, we aim to artificially amplify the signal from $P_{\rm rot}$.

Our initial transformations were normalized to 8 bit integers between [0, 255]. To test how this normalization process affects our training, we rescale the entire transformation to 10× its regular power. Then, we set every pixel with value >255–255, and trained the network on these new normalized transformations. The affect of applying this scaling to the transformations is shown in Figure 7.

Using a training set of $10^6$ examples, we predicted on the associated test set. These results are shown in Figures 6(a) and (l). We find that our new normalization technique decreases the predictive efficiency from 3.81% to 2.83%. While there are numerous rescaling techniques that could be explored, this simple test validates that our initial rescaling technique results in a better performing CNN.

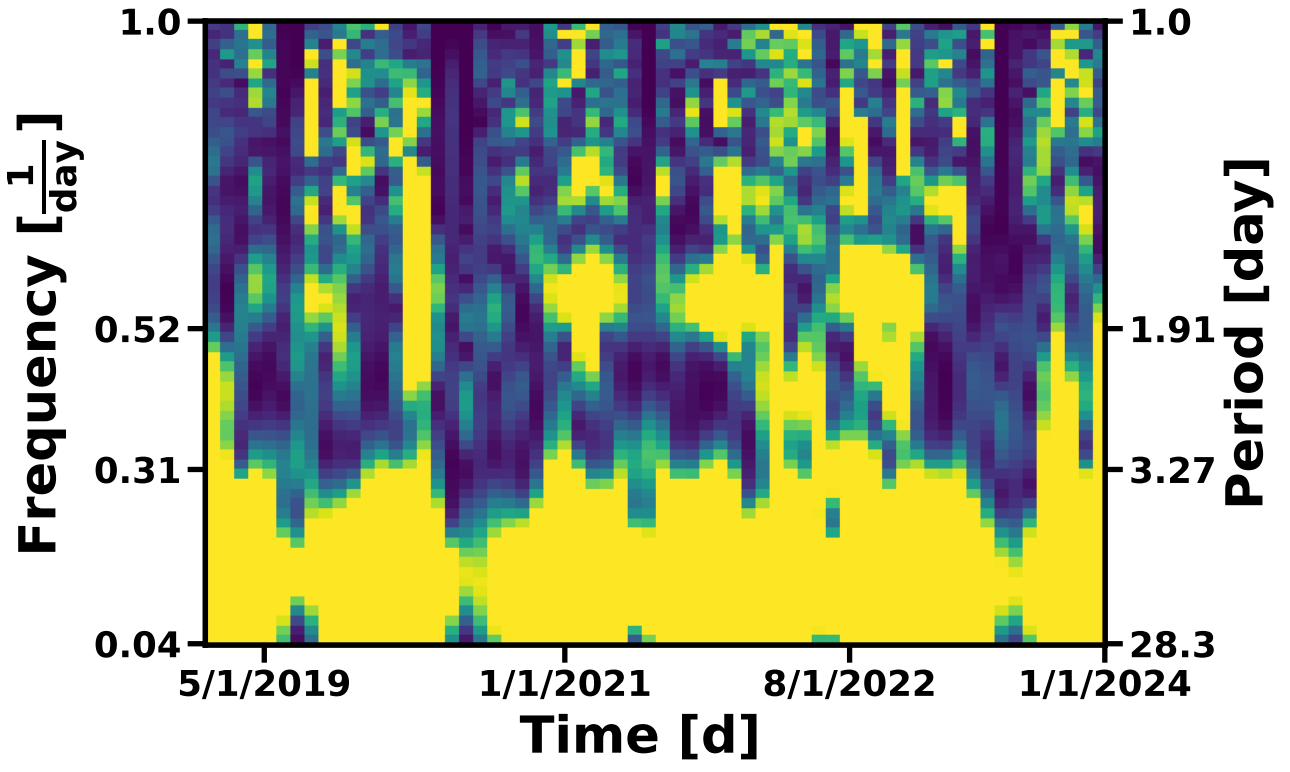

**Figure 7.** The same transformations of ASASSN-V J050623.37-712251.8, the same target presented in Figure 3, but using a different scaling value before being fed into the neural network. While this does suggest that rotational modulation is amplified by this method, it is not as effective as the base procedure, as shown in Figure 6(a) versus 6(l). We use a scaling factor of 10.

## 5. Validation of Rotation Periods

We ran 85,904,442 of the transformed ASAS-SN light curves through our determined "best" network (row 5 of Table 3). We note that the number of transforms we predict on is ∼90% of our full ASAS-SN catalog. This is due to data loss that occurred during our transformation routine performed on HiPerGator. We crossmatch these remaining targets by their IDs with ESA's Gaia Data Release 2 (DR2) catalog. We query the following parameters: parallax, $B_p$, $R_p$, and $G$ magnitudes (D. W. Evans et al. 2018; Gaia Collaboration et al. 2018). Additionally, for our catalog of reported periods, we query Gaia DR3 for stellar luminosity, radii, and spectroscopic rotational broadening (M. Fouesneau et al. 2023). We then used the Gaia `XGBoost` catalog (R. Andrae et al. 2023) to include initial inferred values of the effective temperature ($T_{\rm eff}$), surface gravity (log(g)), and metallicity ([M/H]) of our sample. We note that ∼10% of our sample lacked at least one of the aforementioned parameters, which is propagated into our resulting catalog. We do not speculate as to why these parameters are missing for each individual target.

### 5.1. Predicted Periods and Uncertainties

Our neural network does not supply true errors that can be used to determine accurate predictions. Instead, we must calibrate the errors and define what is a robust detection. As mentioned earlier in this work, we define stars with fractional uncertainties of <25% as robust. Here, we quantitatively justify this cutoff by crossmatching our targets with other works that have presented measured $P_{\rm rot}$. We first crossmatched our predictions to the Gaia DR2-Kepler Input Catalog (KIC) 1″ crossmatch list from the Gaia-Kepler fun website (found at https://gaia-kepler.fun). Then, we used the provided KIC IDs to crossmatch our sample to the catalogs presented in A. R. G. Santos et al. (2019, 2021b), whose $P_{\rm rot}$ were robustly measured based on photometric variability; we identified 41,825 crossmatched stars. We then select targets whose predicted $P_{\rm rot}$ from our neural network was accurate to within 10% of the $P_{\rm rot}$ reported in A. R. G. Santos et al. (2019, 2021b), and compare our sample's fractional uncertainty to the raw $\sigma$-values output from our CNN in Figure 8.

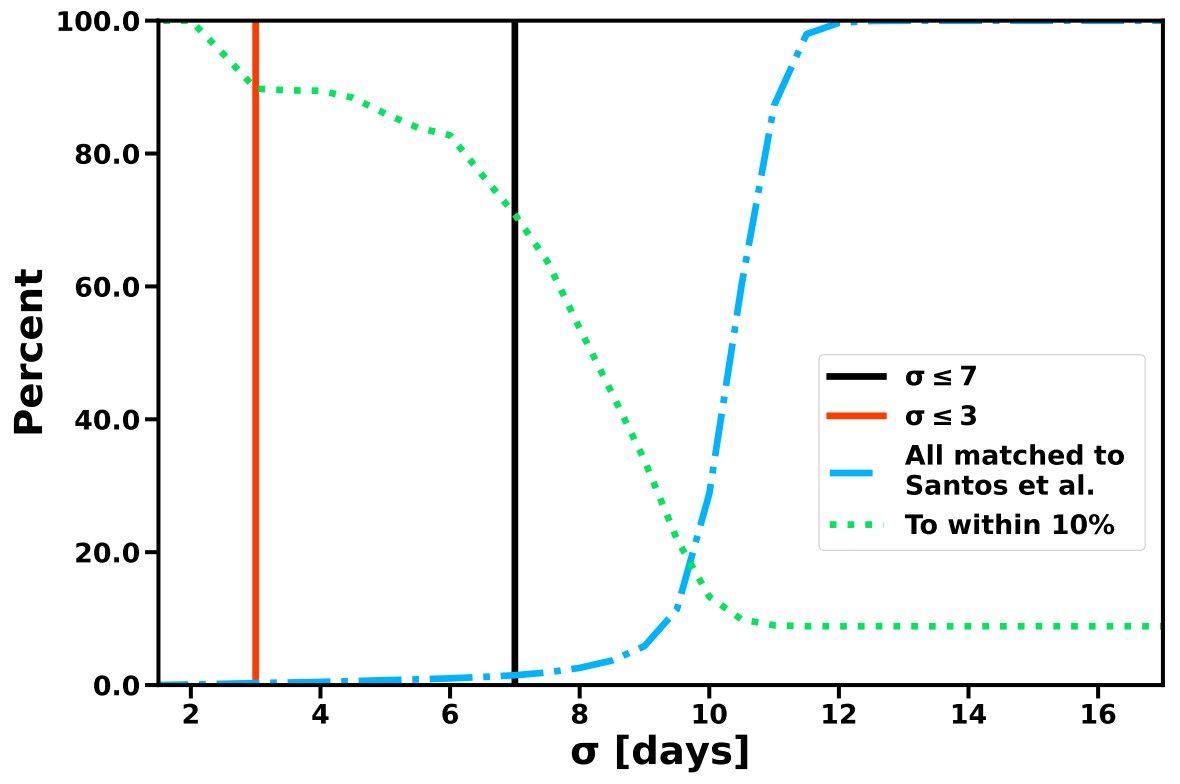

(a) Error

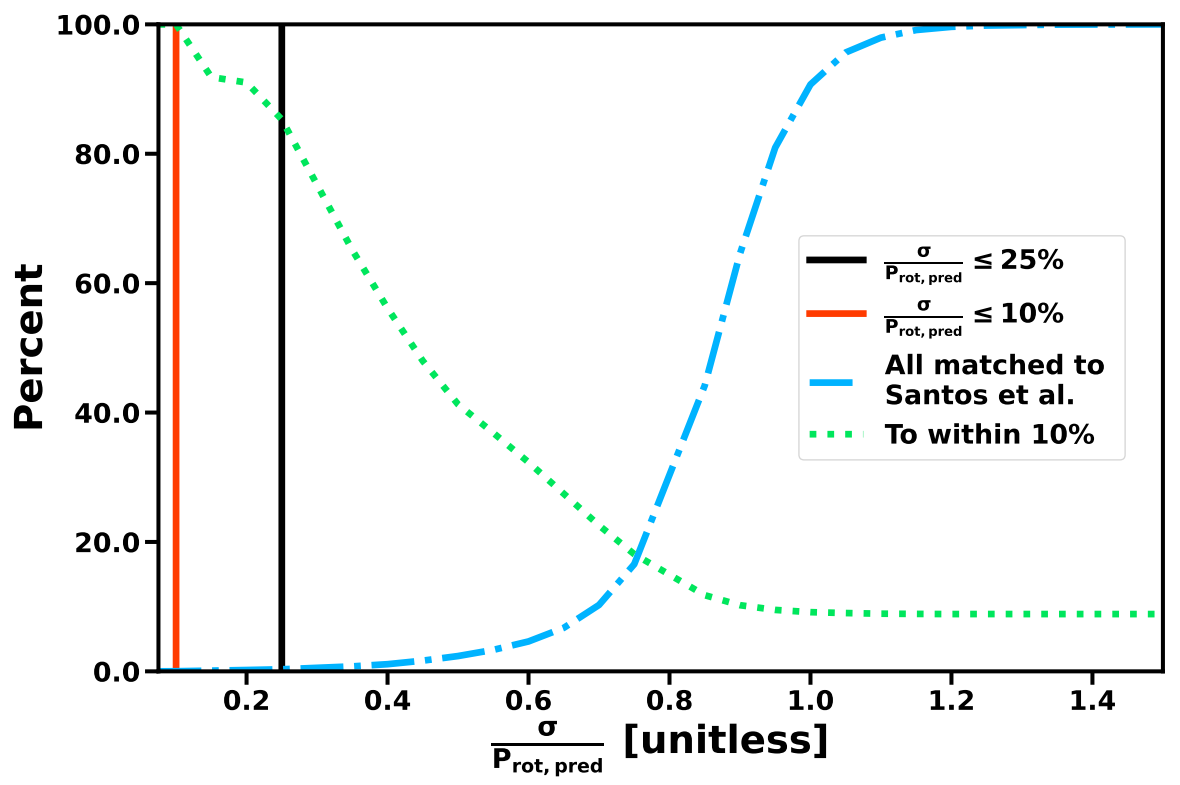

(b) Fractional Uncertainty

**Figure 8.** Justification for our reported "gold" sample being chosen via fractional uncertainty on inferred $P_{\rm rot}$. The top panel is shown with respect to error, while the bottom is in fractional uncertainty-space. A stricter cutoff results in a smaller reported sample, but black cuts indicate period catalogs of similar sizes. By choosing fractional uncertainty as our metric of assessing quality predictions, we maintain a more robust catalog of predictions at all values of $P_{\rm rot}$ while still ensuring our reported catalog is substantial.

Figure 8 demonstrates that increasing the constraint on our error cut improves the reliability of our sample, but reduces the sample size. Furthermore, Figure 8 demonstrates that using the fractional uncertainty provides a balance, where we are still left with many reliably predicted $P_{\rm rot}$ without removing too many stars. We note that using the fractional uncertainty is better at removing spurious short $P_{\rm rot} \leqslant 7$ days. This is likely due to the fractional uncertainty being sensitive to the errors and predictions in conjunction, while an error cutoff is completely independent of our predicted periods. Using a fractional uncertainty cutoff of 25% resulted in a sample of 208,260 stars with reliably predicted $P_{\rm rot}$, defined as our "gold" sample (data model shown in Table 4). The remaining analyses presented in this work only use the $P_{\rm rot}$ from this subset of stars.

We present the spatial distribution and color–magnitude diagram of stars with reported $P_{\rm rot}$ in Figure 9. We demonstrate that we have a high recovery of $P_{\rm rot}$ for giant stars, although it is unlikely these periods are related to rotation (Section 6.2). We detect $P_{\rm rot}$ for FGKM stars and subgiants, and we do not detect $P_{\rm rot}$ for hotter stars, which is to be expected (E. A. Avallone et al. 2022). We further discuss the limitations of our presented $P_{\rm rot}$ by comparing to external catalogs.

**Table 4**
Columns of Our Gold Sample of ASAS-SN Sources with Inferred Periods Found on Zenodo (M. Schochet et al. 2025a)

| Column | Description |
|---|---|
| asas_sn_id | Unique ASAS-SN identifier |
| period | CNN Output $P_{\rm rot}$ |
| sigma | CNN Output $P_{\rm rot}$ error |
| dr3_source_id | Gaia DR3 designation |
| dr2_source_id | Gaia DR2 designation |
| edr3_source_id | Gaia EDR3 designation |
| KIC | Kepler Input Catalog identifier |
| tic_id | TESS Input Catalog identifier |
| ra_rad | Gaia DR2 R.A. (rad) |
| dec_rad | Gaia DR2 decl. (rad) |
| parallax | Gaia DR2 parallax (mas) |
| abs_g_mag | Gaia DR2 $M_G$ (not extinction corrected) |
| Tmag | TESS magnitude |
| phot_g_mean_mag | Gaia DR2 $G$ magnitude |
| phot_bp_mean_mag | Gaia DR2 $BP$ magnitude |
| phot_rp_mean_mag | Gaia DR2 $RP$ magnitude |
| radius_val | Gaia DR2 photometric radius ($R_\odot$) |
| lum_val | Gaia DR2 photometric luminosity ($L_\odot$) |
| ruwe | Gaia DR2 renormalized unit weight error |
| vbroad | Gaia DR3 spectroscopic rotational broadening (km s$^{-1}$) |
| vbroad_error | Gaia DR3 spectroscopic rotational broadening error |
| catwise_w1 | CatWISE W1 (3.4 $\mu$m) magnitude |
| catwise_w2 | CatWISE W2 (4.6 $\mu$m) magnitude |
| mh_xgboost | Gaia `XGBoost` catalog (R. Andrae et al. 2023) [M/H] (dex) |
| teff_xgboost | Gaia `XGBoost` catalog (R. Andrae et al. 2023) effective temperature (K) |
| logg_xgboost | Gaia `XGBoost` catalog (R. Andrae et al. 2023) surface gravity (dex) |
| in_training | "yes" or "no" whether this object is in the training set for this work |
| in_xgboost_training | "True" or "False" whether this object was in the training set for the `XGBoost` parameters |
| simulation_number | what simulation was injected into this object if it was in the training set |

## 5.2. Comparison to Literature Periods

We validate our "gold" sample by crossmatching our predicted $P_{\rm rot}$ with other publicly available catalogs. We aim to validate $P_{\rm rot}$ across a range of spectral types and determine the limitations of our methodology. Fortunately, we had many catalogs to compare to. We select to compare our "gold" sample to the following catalogs: (I) 55,232 stars with measured $P_{\rm rot}$ from Kepler (A. R. G. Santos et al. 2019, 2021b); (II) 40,553 stars with measured $P_{\rm rot}$ from ZTF (Y. L. Lu et al. 2022); (III) 7,245/32,159 stars with measured $P_{\rm rot}$ from TESS/Kepler Bonus light curves, respectively (C24/C25); (IV) 13,504 stars with measured $P_{\rm rot}$ from TESS (R. J. Holcomb et al. 2022); (V) 10,909 stars with measured $P_{\rm rot}$ from TESS (I. L. Colman et al. 2024); and (VI) 53,169 stars with measured $P_{\rm rot}$ from ASAS-SN (VSX, C. T. Christy et al. 2023).[7] We also note that some stars have provided

[7] The ASAS-SN Catalog of Variable Stars can be downloaded at https://asas-sn.osu.edu/variables.

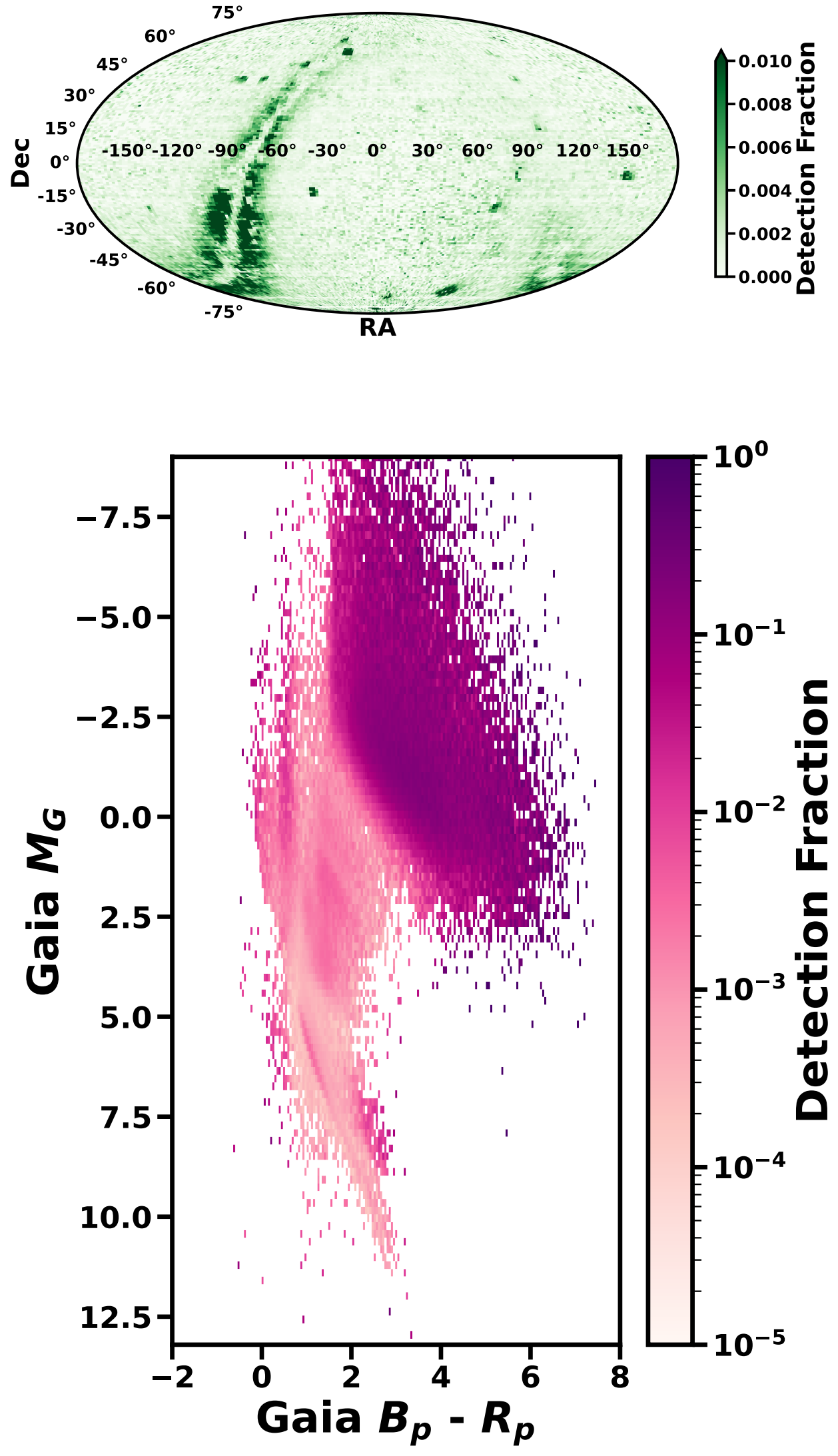

**Figure 9.** The detection fraction of our "gold" sample with inferred $P_{\rm rot}$ spatially distributed across the sky (top) and in color–magnitude space (bottom). We highlight our higher detection fraction along the Galactic plane. In addition, we find that we have a high recovery of periods for stars along the giant branch. This reinforces that our network is most sensitive to bright variable sources.

$P_{\rm rot}$ from several of these catalogs; we highlight this overlap in Appendix C.

Our resulting crossmatch allows us to compare our inferred $P_{\rm rot}$ to tens to thousands of measured $P_{\rm rot}$ from more traditional methods of detection. We compare the measured versus inferred $P_{\rm rot}$ between the aforementioned catalogs and this work in Figure 10. We find that our $P_{\rm rot}$ are in good agreement with those presented in Y. L. Lu et al. (2022), A. R. G. Santos et al. (2019, 2021b), and C24. Meanwhile, we find slight differences in $P_{\rm rot}$ between our work and the remaining catalogs. In particular, we find that our predicted $P_{\rm rot}$ are biased toward longer periods as compared to $P_{\rm rot}$ measured from TESS (R. J. Holcomb et al. 2022; I. L. Colman et al. 2024). Furthermore, as compared to C25, we find an increased scatter slightly offset from a 1:1 agreement in $P_{\rm rot}$, whereas compared to R. J. Holcomb et al. (2022) and I. L. Colman et al. (2024), our $P_{\rm rot}$ are scattered around a 2:1 agreement. These differences are likely due to a combination of our network's inability to predict very short $P_{\rm rot}$ and known issues with detecting longer $P_{\rm rot}$ in TESS observations.

In order to further identify the limitations of our presented method, we compare our predicted $P_{\rm rot}$ from this work against $P_{\rm rot}$ presented in VSX in Figure 11. This highlights how our neural network is performing on a range of different variable types from spotted rotational variables to Cepheids to eclipsing binaries.

We find that our presented $P_{\rm rot}$ are in 1:1 agreement with the majority of spotted rotational variables, semiregular variables, long irregular variables, Cepheid-type variables, young stellar objects, and stars of unknown variable type. On the other hand, our network struggles with RR Lyrae–type, Delta Scuti–type, and eclipsing binary–type variables. We downloaded the TESS light curve for a subset of stars in our catalog that were classified in the VSX and present these light curves phase-folded in Figure 12. We find that we are able to predict $P_{\rm rot}$ reliably for nonrotational variability. However, we are unable to accurately recover exact periods for eclipsing binaries. This is likely due to the limited baseline observed across each eclipse in the ASAS-SN observations. Since ASAS-SN cannot observe continuously through either the primary or secondary eclipses due to its observing strategy, all observations taken during an eclipse will be to zeroth-order randomly distributed within the duration of the eclipse. A more thorough sampling of the eclipses is likely needed for our network to robustly infer eclipsing binary periods (although the survey is not necessarily the prohibitive factor here; see, e.g., D. M. Rowan et al. 2023).

## 6. Observed Trends across Different Subpopulations

We explore observed trends across four different subpopulations: hot stars, giants, cool dwarfs, and subgiants. We divide our sample into these subpopulations based on empirical relationships derived from Gaia $B_p - R_p$ color and $M_G$ magnitude. In particular, we classify hot stars as those with temperatures greater than the Kraft break ($T_{\rm eff} \geqslant$ 6500K; R. P. Kraft 1967), which roughly corresponds to $B_p - R_p < 0.6$ (A. C. Beyer & R. J. White 2024). This is further verified by the color–$T_{\rm eff}$ relationships presented in M. J. Pecaut & E. E. Mamajek (2013) and by E. Mamajek.[8] For the remaining three populations, we divide them based on the following relations. Giants must satisfy the following criteria:

$$M_{\rm G} \leqslant 2(B_{\rm p} - R_{\rm p}) - 4. \tag{8}$$

Whereas cool dwarfs must satisfy the following criteria:

$$M_{\rm G} \geqslant 1.4(B_{\rm p} - R_{\rm p}) + 2.8. \tag{9}$$

Any stars that were not categorized as hot stars, cool main-sequence dwarfs, or giants were categorized as subgiants (Figure 13).

### 6.1. Hot Stars

As hot stars possess negligible surface convection zones, we expect these sources to be less spotted and more rapidly rotating than cooler stars. To verify if our measured $P_{\rm rot}$ are physical, we use the spectroscopic rotational broadening, $v_{\rm broad}$, derived from Gaia. With a combination of $v_{\rm broad}$ and $P_{\rm rot}$, we are able to assess if these stars have reasonable inclinations.

[8] https://www.pas.rochester.edu/~emamajek/EEM_dwarf_UBVIJHK_colors_Teff.txt

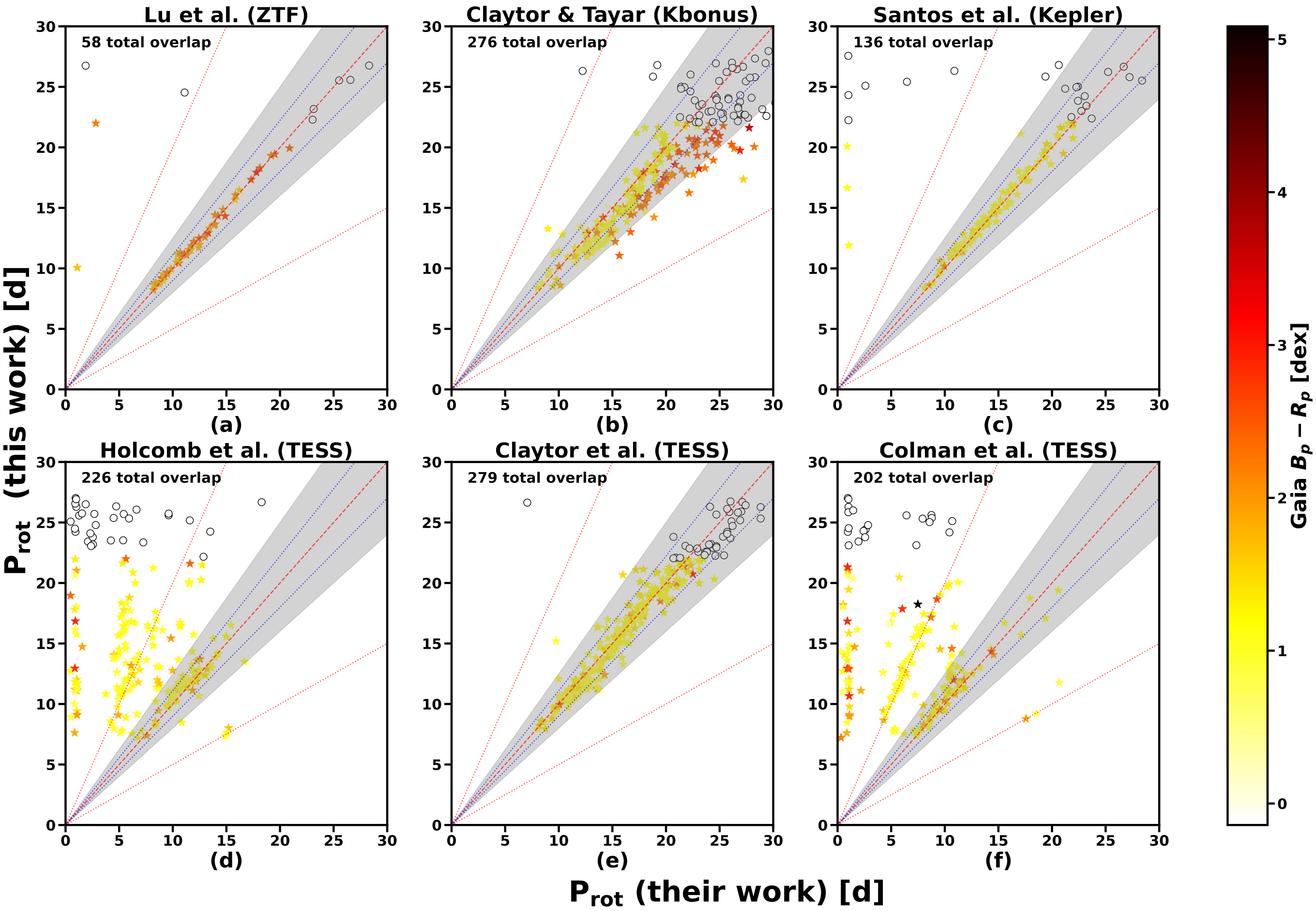

**Figure 10.** Validation of our inferred $P_{\rm rot}$ as compared to several $P_{\rm rot}$ catalogs in the literature. Points are colored by Gaia $B_{\rm p} - R_{\rm p}$, except for the white circles with black borders, which have periods at $P_{\rm rot} = 25 \pm 3$ days (near the sidereal alias). We present crossmatches between: (a) Y. L. Lu et al. (2022), (b) C25, (c) A. R. G. Santos et al. (2019, 2021b), (d) R. J. Holcomb et al. (2022), (e) C24, and (f) I. L. Colman et al. (2024). We present the 1:1 line as the dashed red line and the 90% interval as the blue dotted lines. We plot the 1:2 and 2:1 relationships as the red dotted lines. The shaded region denotes the 25% errors off of the 1:1 trend line between our populations. We find that the majority of our inferred $P_{\rm rot}$ are in agreement with various catalogs, with the exception of the short $P_{\rm rot} < 3$ days stars measured with TESS (R. J. Holcomb et al. 2022; I. L. Colman et al. 2024).

To ensure a robust sample, we discarded any stars with $P_{\rm rot} = 25 \pm 3$ days to ensure we removed any sidereal aliasing that may heavily bias this sample. Additionally, we discarded sources that did not have reported Gaia $v_{\rm broad}$. Finally, we removed stars with $v_{\rm broad}$ measurements below the threshold of accurate broadening detection ($v_{\rm broad} \leqslant 10$ km s$^{-1}$; Y. Frémat et al. 2023). This resulted in a final sample of 146 hot stars.

We computed a $\frac{P_{\rm spec}}{\sin(i)}$ via

$$\frac{P_{\rm spec}}{\sin(i)} = \frac{2\pi R[{\rm km}]}{v_{\rm broad}[\sin(i)\ {\rm km\ s^{-1}}]}, \tag{10}$$

where $R$ is the stellar radius provided by Gaia. We present our inferred periods versus $\frac{P_{\rm spec}}{\sin(i)}$ in Figure 14. We find that nearly all of our inferred periods are spurious for a sample of stars that are likely rapidly rotating with minimal observable photospheric spots. This result suggests that the derived $P_{\rm rot}$ for these hot stars are likely nonphysical, although some of these may be real periods from background stars or smaller companions. Further understanding of why the presented neural network is detecting these $P_{\rm rot}$ is saved for future work.

### 6.2. Oscillating Giants

We aim to determine whether we could identify the oscillatory frequencies of maximum power, $\nu_{\rm max}$, in luminous asteroseismic giants, similarly to C25. To calculate $\nu_{\rm max}$, we adopt the following relationship (see, e.g., T. M. Brown et al. 1991; H. Kjeldsen & T. R. Bedding 1995; W. J. Chaplin et al. 2011)

$$\frac{\nu_{\rm max}}{\nu_{\rm max,\odot}} = f_{\nu_{\rm max}} \frac{g}{g_\odot}\left(\frac{T_{\rm eff}}{T_{\rm eff,\odot}}\right)^{-0.5}. \tag{11}$$

Here, $g$ is the surface gravity of the star, and $f_{\nu_{\rm max}}$ is an empirically derived correction function to ensure the relations scale appropriately for most stars across the possible physical parameter space (for a deeper discussion, see M. H. Pinsonneault et al. 2025). We adopt $T_{\rm eff,\odot} = 5772$K, $g_\odot = 10^{4.438}$ (both A. Prša et al. 2016), $\nu_{\rm max,\odot} = 3076\ \mu$Hz, and $f_{\nu_{\rm max}} = 1$ (both M. H. Pinsonneault et al. 2025). We derive $T_{\rm eff}$ and $g$ from the catalog of R. Andrae et al. (2023), which provides machine learning derived parameters (`XGBoost` algorithm; T. Chen & C. Guestrin 2016) trained on Sloan Digital Sky Survey (SDSS) APOGEE DR17 (Abdurro'uf et al. 2022). Approximately ~88% of our giant sample have available derived $T_{\rm eff}$ and log($g$) from this catalog. We use these derived parameters to then calculate $\nu_{\rm max}$. We convert these $\nu_{\rm max}$ into a period such that we could compare our inferred $P_{\rm rot}$ to the oscillatory period of maximum power (Figure 15).

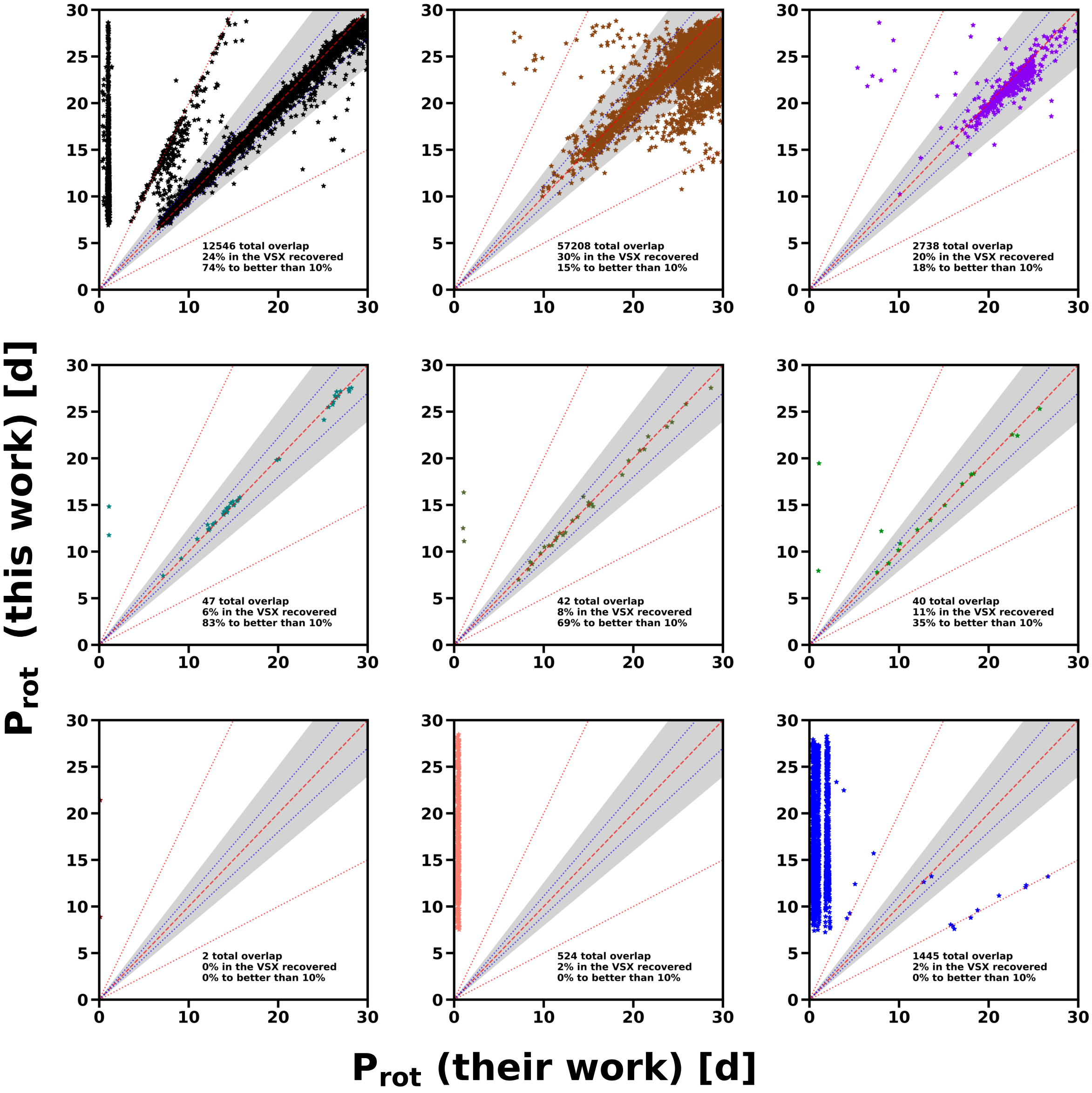

**Figure 11.** A comparison between our inferred $P_{\rm rot}$ and those presented in C. T. Christy et al. (2023). We separate our sample by the assigned variable type from VSX. We show the comparison for spotted rotational (black, top left), semiregular (brown, top center), long irregular (purple, top right), Cepheid-type (teal, center left), unknown variable type (dark green, true center), young stellar objects (light green, center right), Delta Scuti–type (maroon, bottom left), RR Lyrae–type (orange, bottom center), and eclipsing binaries (blue, bottom right). The shaded regions here again denote the 25% fractional uncertainty range at different value of $P_{\rm rot}$. These plots demonstrate an extremely low recovery of the true $P_{\rm rot}$ for short-term variables like RR Lyrae–type and Delta Scuti–type stars. In addition, our network is able to recover $P_{\rm rot}$ for spotted, semiregular, and irregular variables well. The clumping of stars at $P_{\rm rot} \leqslant 3$ days are due to the limitations of our network.

We find that our inferred periods are closely correlated to the derived $\nu_{\rm max}$. We continue to find a significant fraction of stars with inferred $P_{\rm rot} = 25 \pm 3$ days, corresponding to the sidereal alias. We find that $P_{\rm rot} < \nu^{-1}_{\rm max,XGBoost}$ are on average redder than the population with inferred $P_{\rm rot} > \nu^{-1}_{\rm max,XGBoost}$. Note that 76.5% of our remaining giant sample possess $P_{\rm rot} > \nu^{-1}_{\rm max,XGBoost}$.

We aim to determine whether the offset between inferred $P_{\rm rot}$ and derived $\nu_{\rm max}$ is due to our neural network or uncertainties from the `XGBoost` parameters. As such, we crossmatched our giant sample to the APOKASC-3 catalog (M. H. Pinsonneault et al. 2025), which has well-calibrated asteroseismic parameters for a sample of stars contained in both SDSS APOGEE and Kepler. We find 146 stars in our sample that overlap with the APOKASC-3 catalog (Figure 16). We demonstrate that our inferred $P_{\rm rot}$ are not offset from the APOKASC-3 $\nu_{\rm max}$, which leads us to conclude that our inferred $P_{\rm rot}$ for giants are generally consistent with $\nu_{\rm max}$, and it is likely the `XGBoost` parameters are slightly offset from the true values. However, we caution that additional validation is required should future work choose to use our our neural network to derive $\nu_{\rm max}$.

### *6.3. Cool Dwarfs*

We aim to verify that our catalog of inferred $P_{\rm rot}$ for cool stars displays the commonly identified "intermediate period

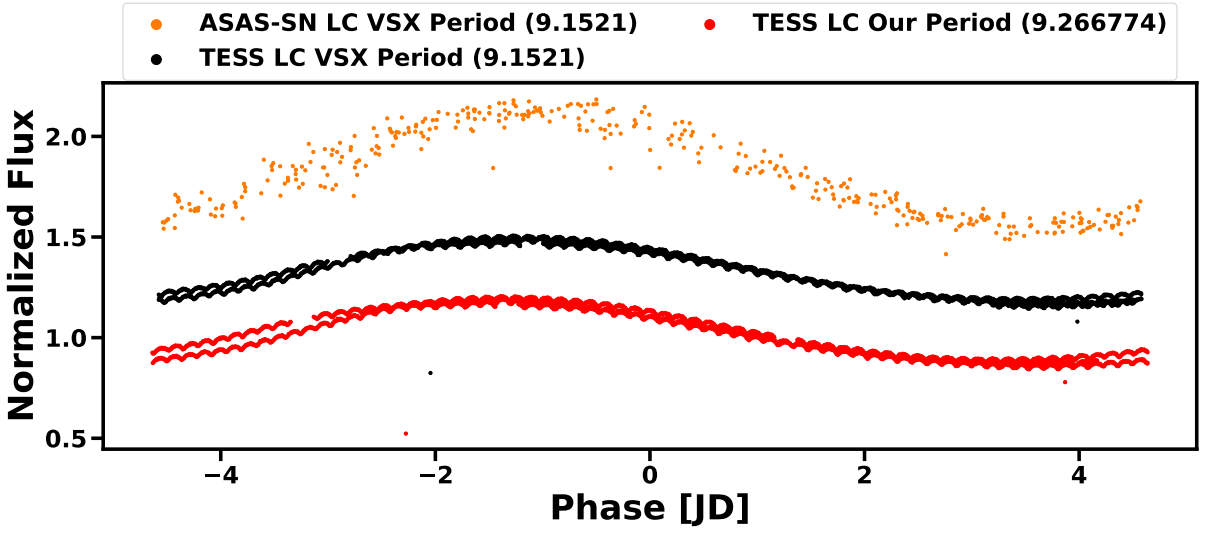

(a) TIC 201574467 (W Virginis-type Cepheid variable)

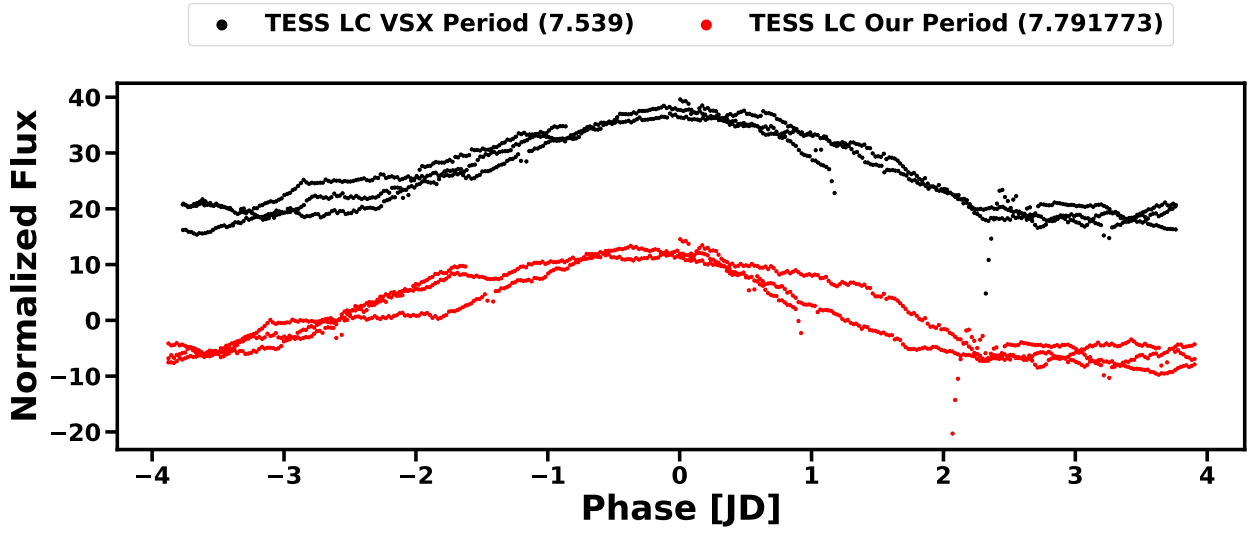

(b) TIC 436243677 (Young stellar object)

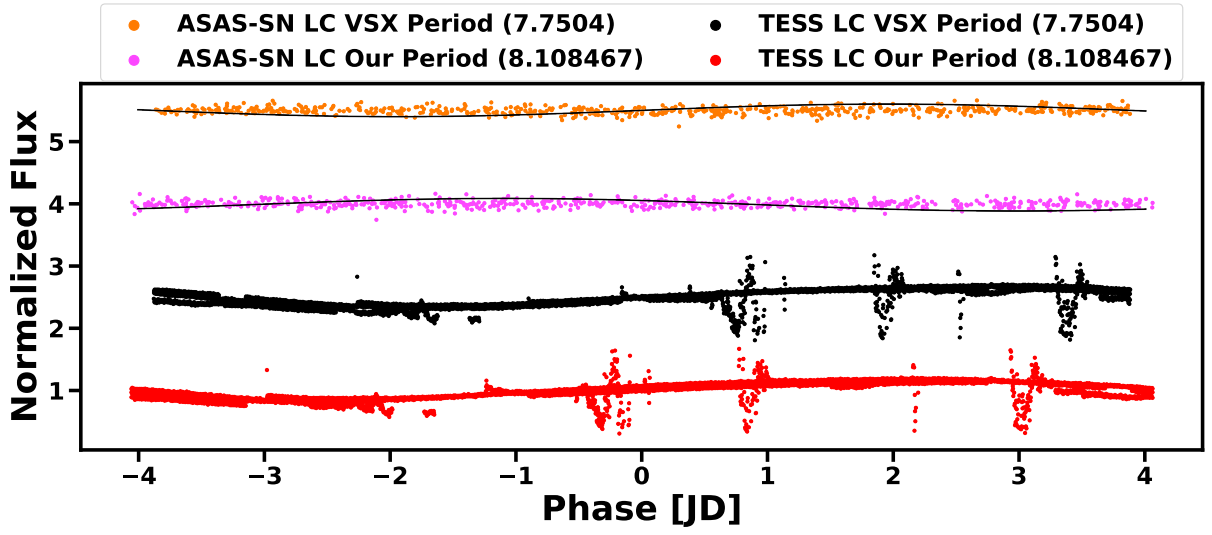

(c) TIC 6129951 (Spotted rotational variable)

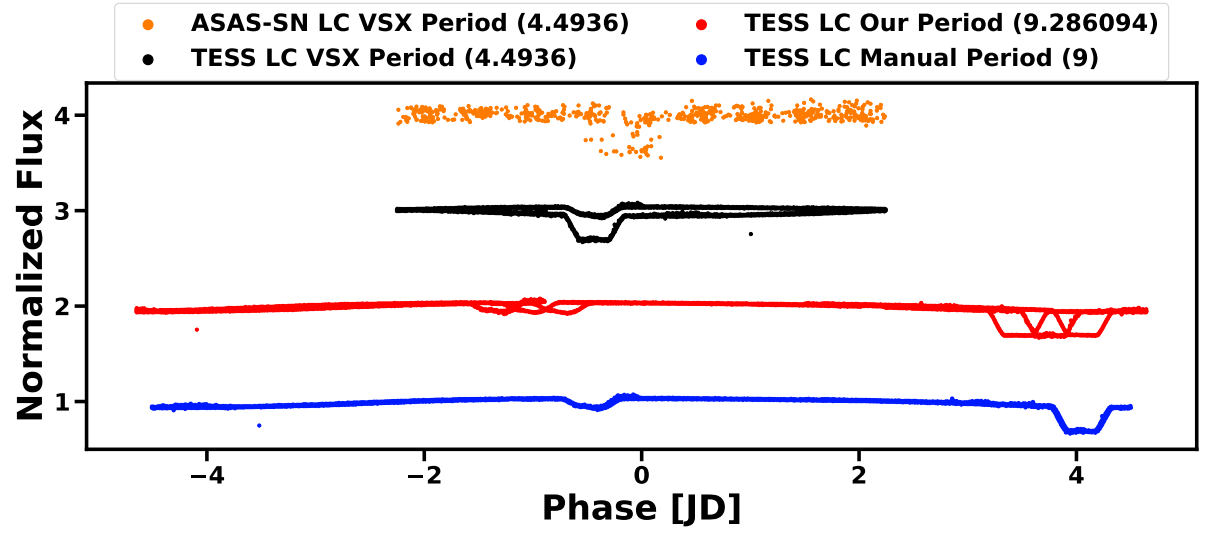

(d) TIC 397585242 (Algol-type eclipsing binary)

**Figure 12.** TESS and ASAS-SN phase-folded light curves for a (a) Cepheid, (b) young stellar object, (c) spotted rotational variable, and (d) eclipsing binary, as categorized by the VSX. Our network is able to recover nonrotational variables (panels (a) and (b)). Our network is also able to recover spot-driven rotational variability (panel (c)), though there are offsets from previous work. We additionally highlight the network's ability to recover the true period of eclipsing binaries (panel (d)).

gap" that has been seen across several different studies (e.g., A. McQuillan et al. 2013; J. R. A. Davenport 2017; J. R. A. Davenport & K. R. Covey 2018; T. Reinhold et al. 2019; J. L. Curtis et al. 2020; F. Spada & A. C. Lanzafame 2020; T. A. Gordon et al. 2021; Y. L. Lu et al. 2022).

Before we proceeded to vet these periods, we wanted to ensure that the reported periods are representative of the periods found in nature. We plot the Gaia photometric magnitude distribution of our dwarfs and oscillating giants in Figure 17. This figure highlights the fact that the majority of our inferred periods are from bright sources; however, there are a substantial number of stars out to much dimmer magnitudes. In the end, to demonstrate the highest-fidelity catalog of cool dwarf periods, we decided to cut out any dwarfs dimmer than a Gaia photometric magnitude of 15. This removed a significant fraction of stars near the sidereal period (∼27 days), and reduced our cool dwarf sample from 38,860 to 29,473 stars.

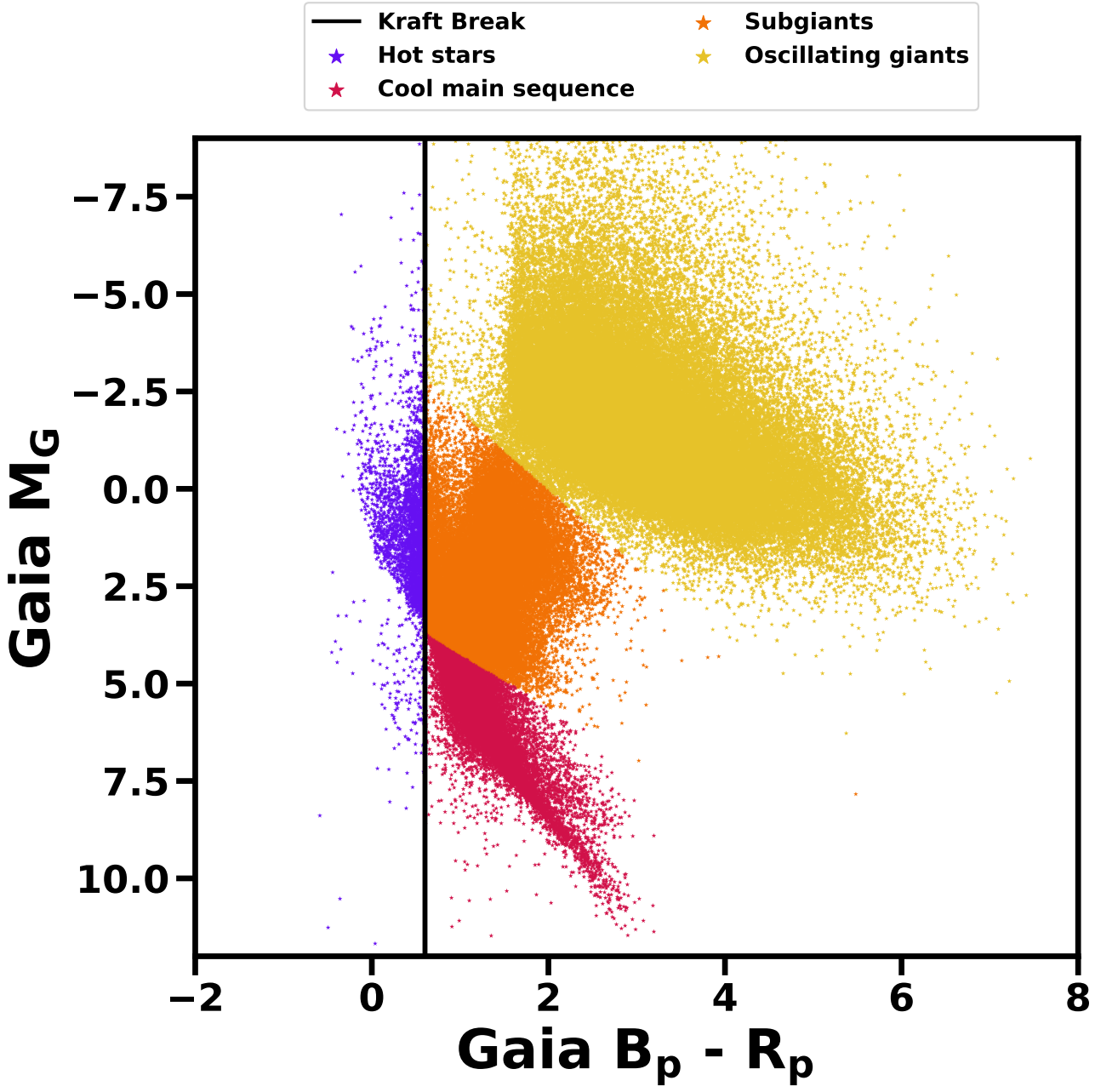

**Figure 13.** An H-R diagram of our "gold" sample (208,260 inferred $P_{rot}$) from our neural network. We highlight our empirical relationships to distinguish between four subpopulations: hot stars (purple), oscillating giants (yellow), cool main-sequence stars (red), and subgiants (orange).

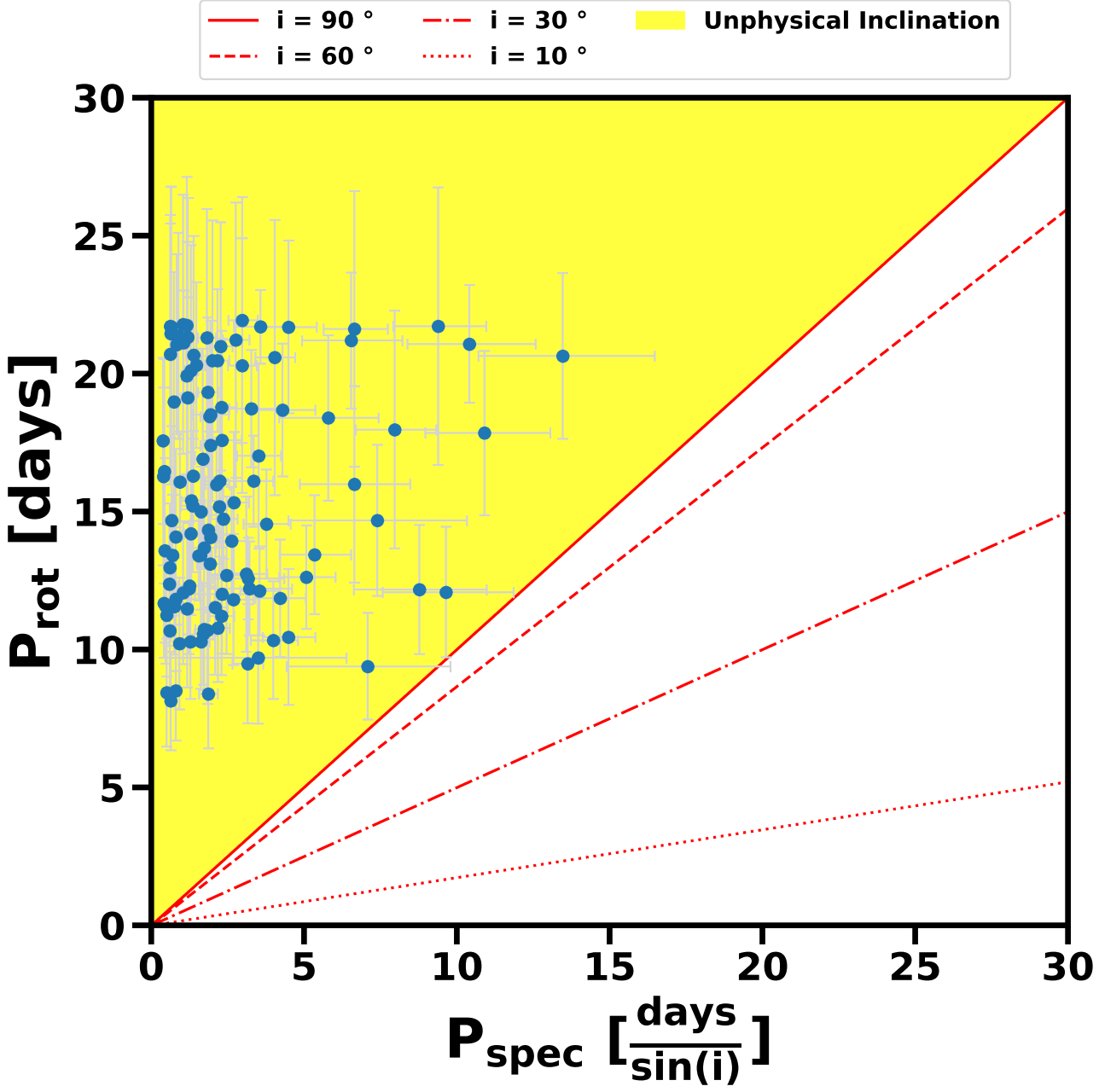

**Figure 14.** A comparison between our inferred $P_{rot}$ and the computed $\frac{P_{spec}}{\sin(i)}$ for our hot star subpopulation. The red lines indicate curves of constant inclination. The shaded region (yellow) denotes nonphysical inclinations. Because all of our inferred $P_{rot}$ for hot stars lie within this nonphysical region, we conclude our hot star periods are likely spurious.

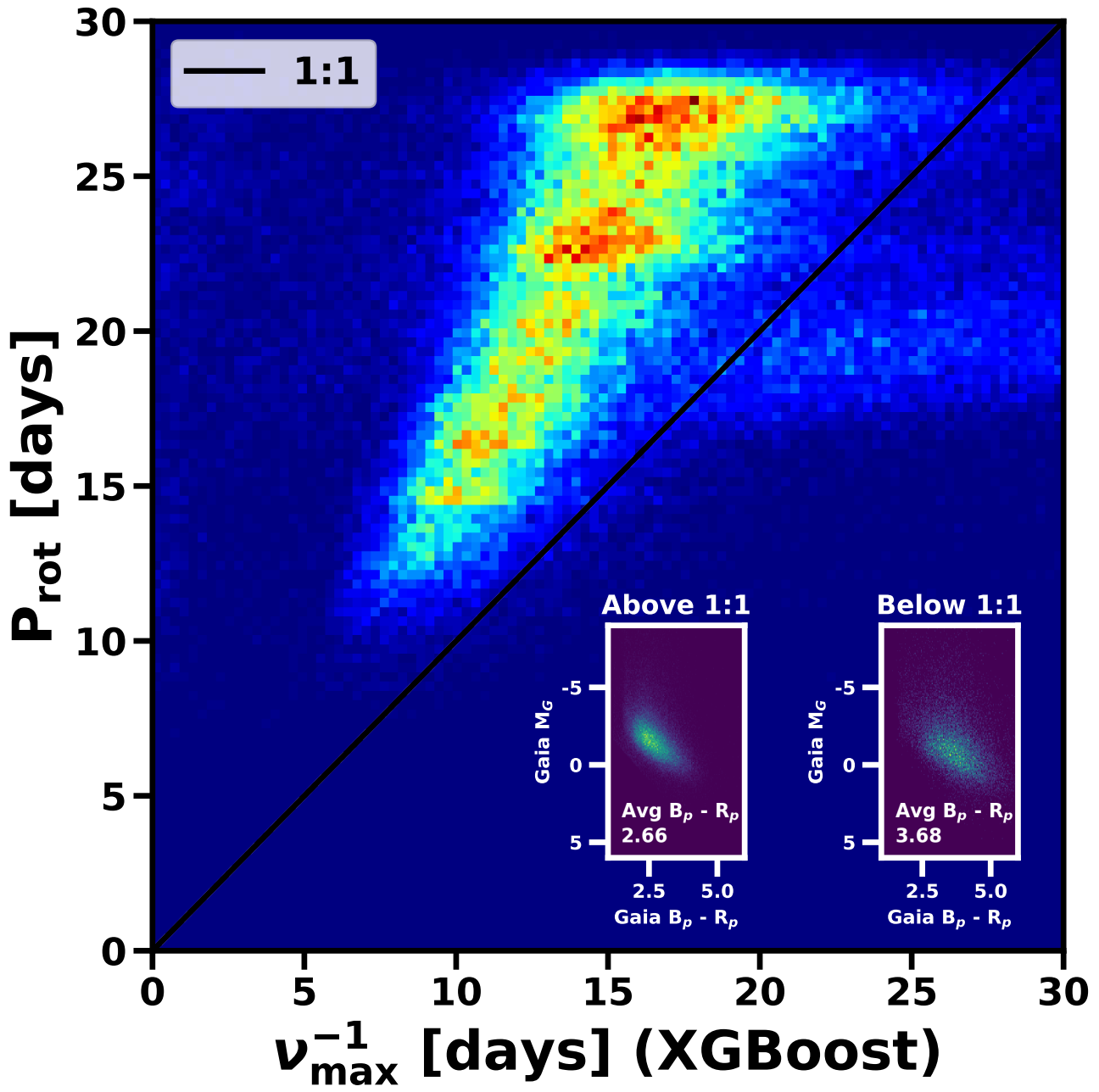

**Figure 15.** A comparison between our inferred $P_{\rm rot}$ and the computed $\nu_{\rm max}^{-1}$ from Gaia XGBoost parameters. While our predictions trace the expected trend, our catalog is systematically offset from the estimated $\nu_{\rm max}^{-1}$ by ∼65%. The inset H-R diagrams demonstrate that stars falling above the 1:1 line (73,301 total, 67.7%) are systematically bluer than stars that fall below the 1:1 line (35,060 total, 32.4%).

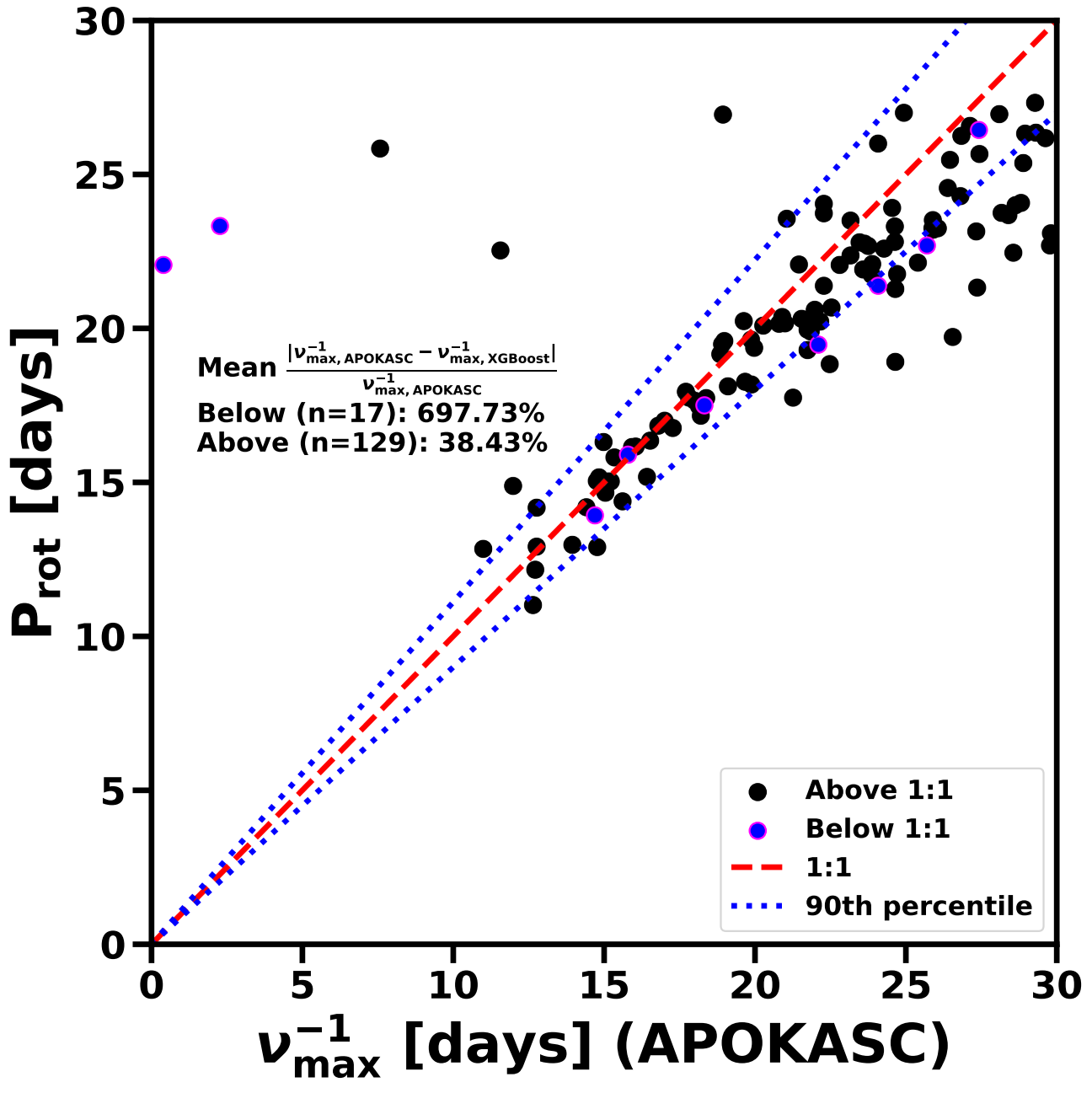

**Figure 16.** A comparison of $\nu_{\rm max}^{-1}$ computed from the APOKASC-3 catalog to our inferred $P_{\rm rot}$. We demonstrate that the majority of our matched $P_{\rm rot}$ are in agreement with the asteroseismic-derived $\nu_{\rm max}^{-1}$ values from APOKASC-3. Thus, we demonstrate that our inferred $P_{\rm rot}$ for giants are likely astrophysical and are correlated with observed oscillatory periods.

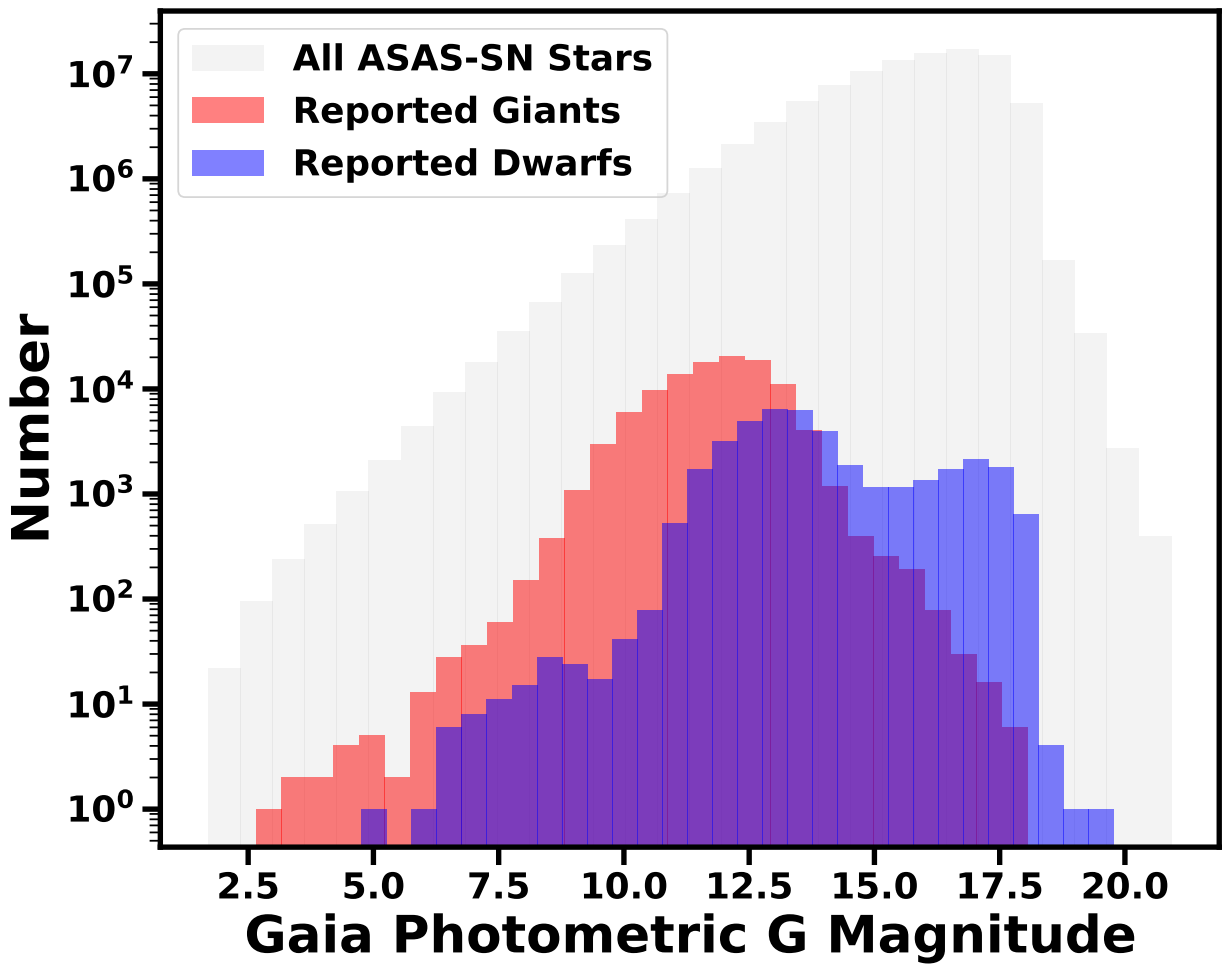

**Figure 17.** A 2D histogram of the Gaia photometric magnitudes for our oscillating giant and dwarf subsamples compared to the full ASAS-SN catalog. This highlights both that the majority of our oscillating giant periods are from the brightest stars, and that our inferred periods for dwarfs trail off to much dimmer magnitudes. This suggests that our inferred periods for the dimmest stars—especially for cool dwarfs—may be spurious.

We use the "period gap" boundaries defined in T. A. Gordon et al. (2021): each line is defined as

$$P_{\rm upper} = A(G - G_{\rm RP} - x_0) + B(G - G_{\rm RP} - x_0)^{1/2}. \quad (12)$$

Parameters $A$ and $B$ have units of days. For the upper edge, $A = 68.2277$, $B = -43.7301$, and $x_0 = -0.0653$; for the lower edge, $A = 34.0405$, $B = -2.6183$, and $x_0 = 0.3510$. We crossmatch the $P_{\rm rot}$ presented in A. R. G. Santos et al. (2019, 2021b) with Gaia to compare to our sample. We plot our inferred periods alongside the Santos periods in Figure 18. Overall, we find that our inferred cool dwarf periods line up quite well with the periods from the Santos catalogs, except for the clumping of stars at $P_{\rm rot} = 25 \pm 3$, which we attribute to sidereal aliasing. We note that although the cut to remove the dimmest stars did reduce the number of periods found at this alias, the alias is still difficult to entirely remove from our sample; periods close to this value should be treated with caution. We also note that the period gap is more accurately an underdensity in period–color space, and that both our predictions and those of the Santos catalogs do contain a few predictions within the gap boundaries.

### 6.4. Subgiants

We compare our subgiant population to the catalog of E. M. Leiner et al. (2022; Figure 19) to determine if our inferred $P_{\rm rot}$ are physical. We find there are 574 stars overlapping between both catalogs and that our network infers an accurate $P_{\rm rot}$ for the majority of stars matched, all of which are anomalous sub-subgiants or RSCVns. We are able to recover 467 (81.35%) of the E. M. Leiner et al. (2022) catalog to within 10% of their period as reported in the American Association of Variable Star Observers Variable Star Index. Overall, we find the average percent difference across these 467 stars is ⩽1.9%, suggesting our network excels at inferring true $P_{\rm rot}$ for unusually active and rapidly rotating subgiants.

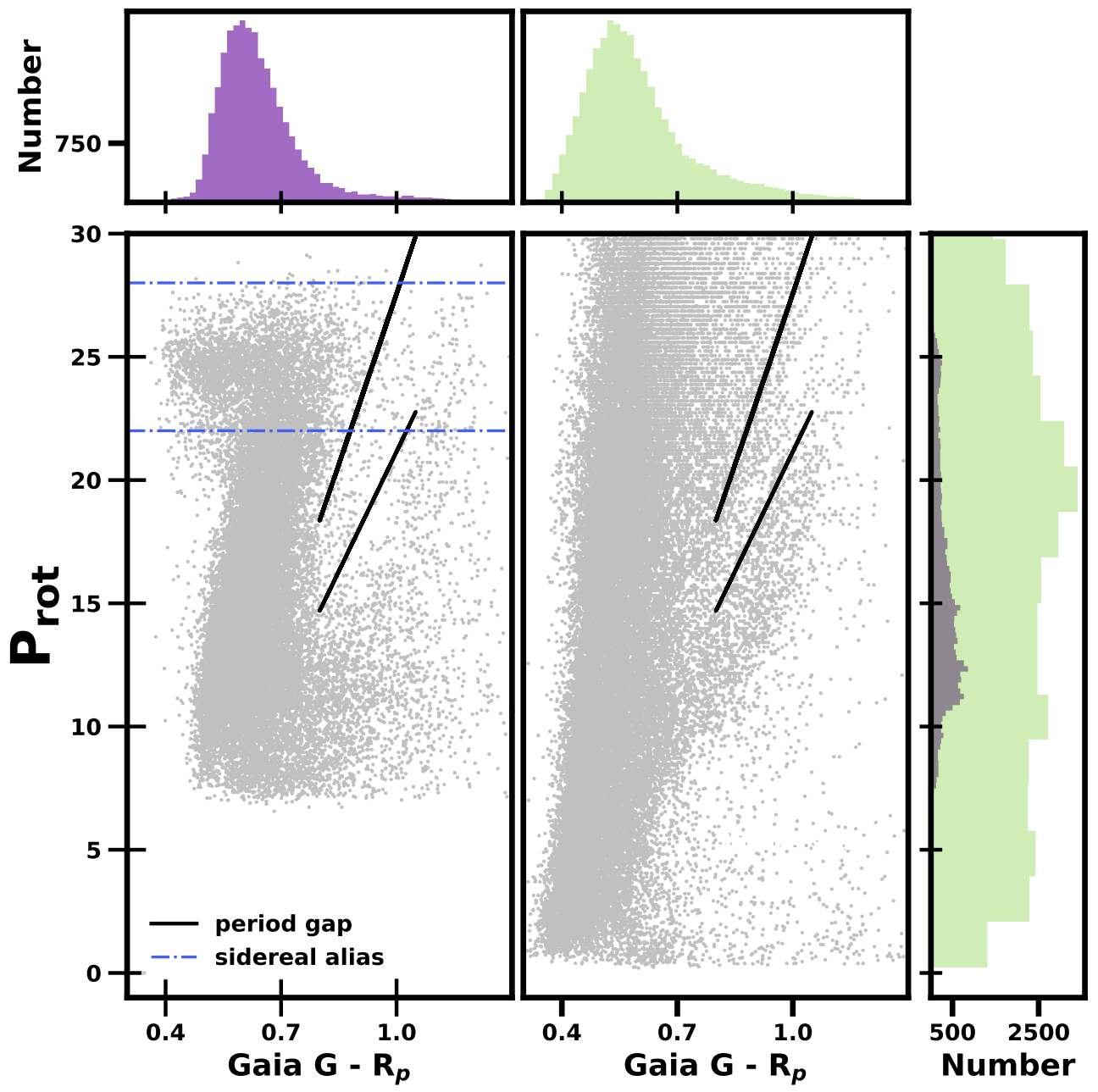

**Figure 18.** Our inferred $P_{\rm rot}$ of our brightest cool dwarfs (left) compared to those from the A. R. G. Santos et al. (2019, 2021b) catalogs (right). We mark the edges of the "period gap" (T. A. Gordon et al. 2021) in both panels with the black lines. Our sample shows a similar distribution compared to Kepler, with the exception the large clumping of stars at $P_{\rm rot} = 25 \pm 3$ days (sidereal alias).

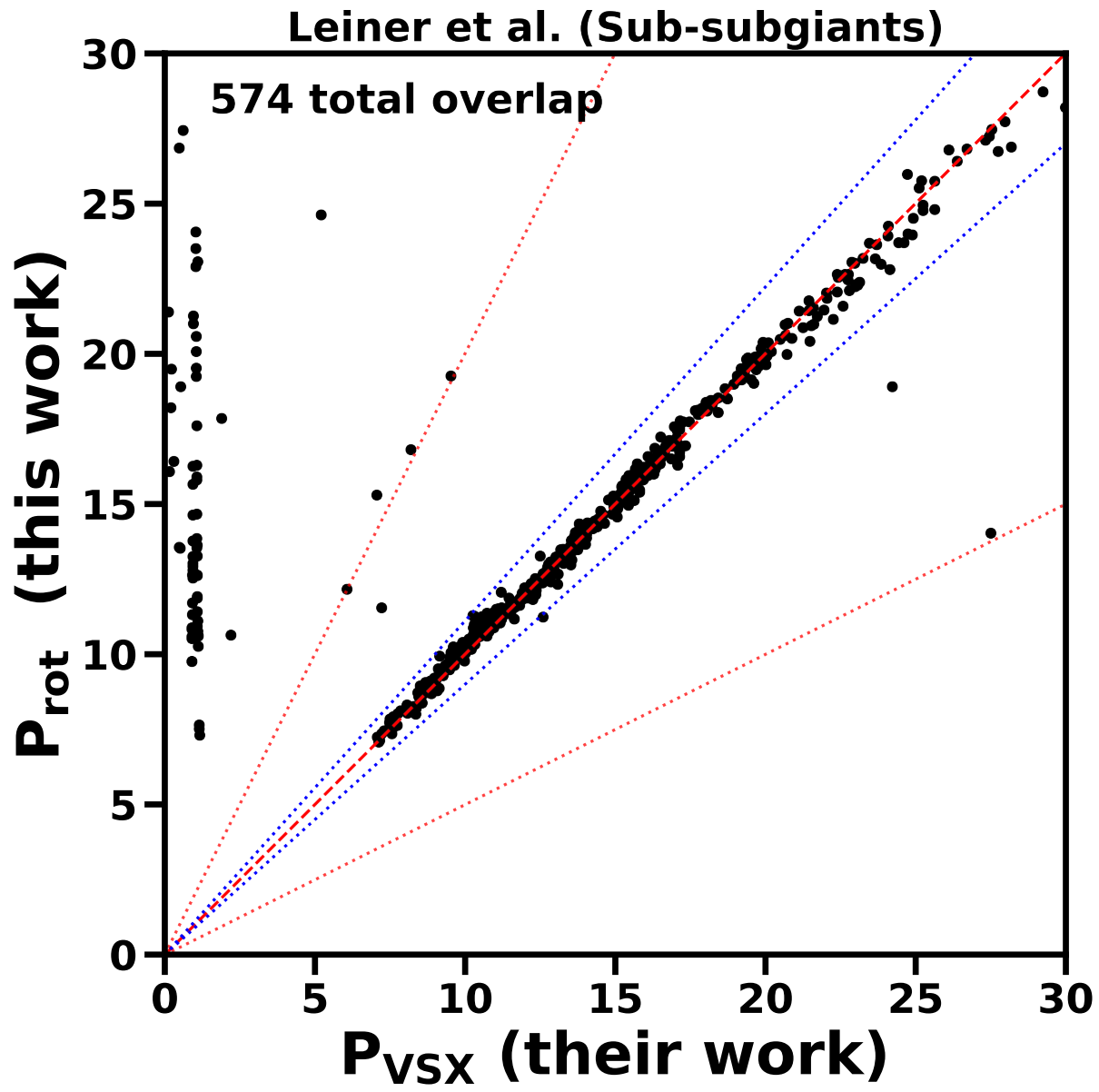

**Figure 19.** A comparison between our inferred $P_{\rm rot}$ and the $P_{\rm rot}$ taken from the American Association of Variable Star Observers Variable Star Index for stars in E. M. Leiner et al. (2022). We find that the majority of our $P_{\rm rot}$ are within ~10% agreement with those presented in E. M. Leiner et al. (2022). Several stars at short periods ($P_{\rm rot} < 5$ days) are in disagreement, which is unsurprising due to the limitations of our neural network.

## 7. Conclusion

We present a new neural network trained to infer variability periods for stars observed by ASAS-SN. We trained our neural network on simulated light curves (Z. R. Claytor et al. 2022). We performed various tests to validate the robustness of our neural network, including changing our training set and using different network architectures. From the full ASAS-SN catalog, we are able to infer 208,260 periods. Note that 196,610 of these stars were matched with Gaia counterparts and explored in Section 6. We provide 108,361 stars (55.11%) categorized as giants, 38,860 stars (19.76%) categorized as cool dwarfs, 45,909 (23.35%) categorized as subgiants, and 3480 (1.77%) categorized as hot stars. Within those predictions, we validate a subset of those periods and report the following:

1. We infer 3480 periods for hot stars (Gaia $B_p - R_p < 0.6$) that we determine are likely spurious (Section 6.1).
2. We infer 108,361 periods from luminous giant stars. We find that these periods correspond to expected oscillatory frequencies derived from asteroseismic relations (Section 6.2).
3. We infer 29,473 periods for bright cool main-sequence dwarfs, 26,593 of which are new in ASAS-SN. By crossmatching our rotation periods, we find our catalog follows known trends in period–color space, including a dearth of measured $P_{\rm rot}$ in the intermediate period gap (Section 6.3).
4. We infer 45,909 periods for a mixed sample of stars that includes subgiants, active RSCVn-type systems, and anomalous sub-subgiants, which are in agreement with previous works (Section 6.4).

Our results highlight the strengths and limitations of our network architecture. We are reliably able to infer periods driven by spotted photometric variability and asteroseismic oscillations. On the other hand, our architecture struggles to detect confident rotation periods for hot stars, stars with $P_{\rm rot} < 7$ days, and stars with periods near the sidereal month (~27 days). Detecting rotation periods in hot stars is understandably difficult due to limited observable spots, while the sidereal aliasing for variability periods in ASAS-SN observations has been well documented (T. Jayasinghe et al. 2020). For the recovery of the shortest period variables, we can likely attribute the failed detections to intrinsic network architecture problems (as can be seen in the networks of C22, C24, and C25). Regardless, we successfully demonstrate that we are able to recover $P_{\rm rot}$ from sparsely sampled ground-based observations, and that our network has an overall periodicity recovery rate of 0.2%.

We believe this work can serve as the framework for applying deep-learning methods to other ground-based surveys like LSST as well as space-based observatories with complex sampling such as the Nancy Grace Roman mission. Taking the conclusions of this study at face value, we can begin to contextualize our work within the scope of these surveys.

LSST is poised to observe more than 10 billion stars over its 10 yr baseline. Under the assumption that a similar network to ours with 1/10 the efficacy was run on the 10 yr LSST data, we expect LSST to yield of the order of 500,000 stellar rotation periods. However, this is a wildly conservative estimate. Early work on profiling the potential of Rubin suggested that the 10 yr baseline could optimistically measure rotation periods at around 70%–80% completeness for cool stars ($T_{\rm eff} \leqslant 4500$ K) brighter than an $r$-magnitude of 23 (S. L. Hawley et al. 2016). As a result, with a more optimistic outlook on the potential of LSST science, we expect that a similar network to ours could feasibly infer millions of stellar rotation periods.

Meanwhile, the upcoming Nancy Grace Roman mission is expected to launch before mid-2027. One of its core surveys is the Galactic Bulge Time Domain Survey, which should observe six bulge fields continuously over six separate 72 day sectors. While the sampling rate of Roman is still under consideration, it is already expected that the cadence will be more complicated and irregular than other space-based photometric missions like Kepler or TESS. Preliminary modeling of the Roman stellar rotation period yield has placed expectations in the range of 0.05%, resulting in an expected yield of around 100,000 rotation periods (Z. Claytor & J. Tayar 2024).

## Acknowledgments

We acknowledge University of Florida Research Computing for providing computational resources and support that have contributed to the research results reported in this publication (https://www.rc.ufl.edu).

M.E.S., P.P., Z.C., and J.T. acknowledge support from Roman Preparatory grant 80NSSC24K0081. A.D.F. acknowledges funding from NASA through the NASA Hubble Fellowship grant HST-HF2-51530.001-A awarded by STScI. We would like to thank our anonymous referee for keen insights that helped us improve the scope and results of this work. We would also like to thank Rae Holcomb, Isabel Colman, Collin Christy, and Dan Hey for sharing data products used in this project. M.E.S would like to thank Sophie Clark, Sonja Walker, Tessa Frincke, Noah Vowell, Jay Strader, Sydney Petz, and Darryl Seligman for helpful discussions and comments that improved this work. M.E.S. and P.P. would also like to acknowledge Jack Moreland for highlighting that the results of this work are "20+ years in the making" (when accounting for computation time). We thank Chris Kochanek, Ben Shappee, and Kris Stanek for helpful comments on early drafts that improved the quality of this manuscript.

We would like to extend a hearty acknowledgment to the ASAS-SN team at the University of Hawaii for their advice, assistance, and help with the transfer of the complete ASAS-SN catalog needed during the creation of this manuscript. In particular, we would like to thank Kyle Hart, Ben Shappee, Curt Dodds, and Joel Ong for all of their help.

We thank Las Cumbres Observatory and its staff for their continued support of ASAS-SN. ASAS-SN is funded by Gordon and Betty Moore Foundation grants GBMF5490 and GBMF10501 and the Alfred P. Sloan Foundation grant G2021-14192.

This paper includes data collected with the TESS mission, obtained from the MAST data archive at the Space Telescope Science Institute (STScI). Funding for the TESS mission is provided by the NASA Explorer Program. STScI is operated by the Association of Universities for Research in Astronomy, Inc., under NASA contract NAS 5–26555.

This work has made use of data from the European Space Agency (ESA) mission Gaia (https://www.cosmos.esa.int/gaia), processed by the Gaia Data Processing and Analysis Consortium (DPAC, https://www.cosmos.esa.int/web/gaia/dpac/consortium). Funding for the DPAC has been provided by national institutions, in particular the institutions participating in the Gaia Multilateral Agreement.

This research has made use of the SIMBAD database, operated at CDS, Strasbourg, France. SIMBAD may be accessed at https://simbad.u-strasbg.fr/simbad.

This research made use of the crossmatch service provided by CDS, Strasbourg. XMatch may be accessed at http://cdsxmatch.u-strasbg.fr/.

This research has made use of the VizieR catalog access tool, CDS, Strasbourg, France. VizieR may be accessed at https://vizier.cds.unistra.fr/viz-bin/VizieR.

This research has made use of the Catalog Archive Server Jobs System (CasJobs). CasJobs was originally developed by the Johns Hopkins University / Sloan Digital Sky Survey (JHU/SDSS) team, and the original website can be accessed at http://casjobs.sdss.org/CasJobs.

This work made use of the gaia-kepler.fun crossmatch database created by Megan Bedell. The original website can be accessed at http://gaia-kepler.fun/.

We acknowledge that for thousands of years, the area now comprising the state of Florida has been, and continues to be, home to many Native Nations. We further recognize that the main campus of the University of Florida is located on the ancestral territory of the Potano and of the Seminole peoples. The Potano, of Timucua affiliation, lived here in the Alachua region from before European arrival until the destruction of their towns in the early 1700s. The Seminole, also known as the Alachua Seminole, established towns here shortly after but were forced from the land as a result of a series of wars with the United States known as the Seminole Wars. We, the authors, acknowledge our obligation to honor the past, present, and future Native residents and cultures of Florida.

*Software*: `pyasassn` (K. Hart et al. 2023) `butterpy` (Z. R. Claytor et al. 2022; Z. Claytor et al. 2024a), `Lightkurve` (Lightkurve Collaboration et al. 2024), `AstroPy` (Astropy Collaboration et al. 2013, 2018, 2022), `astroquery` (A. Ginsburg et al. 2019), `iPython` (F. Perez & B. E. Granger 2007), `Matplotlib` (J. D. Hunter 2007), `NumPy` (C. R. Harris et al. 2020), `Pandas` (W. McKinney 2010), `polars` (R. Vink et al. 2025), `PyTorch` (A. Paszke et al. 2019), `SciPy` (P. Virtanen et al. 2020).

## Author Contributions

M.E.S., P.P., and J.T. led the collection of ASAS-SN data. M.E.S., P.P., Z.R.C., and J.T. led the development of the CNN architecture and the determination of hyperparameters. M.E.S. led the data preparation, tests on variations of the CNN, the curation of external data sources, and the scientific analysis of our reported catalog. Z.R.C. and A.D.F. contributed extensively to the methodology of this work, and also provided substantial manuscript text and review.

## Data Availability

We host the majority of data products and catalogs created to accomplish this work on Zenodo (M. Schochet et al. 2025a) at doi:10.5281/zenodo.15848601. In particular, we host the following files:

1. `allgaiadata.parq`—Gaia DR2 colors, magnitudes, and parallaxes crossmatched by ID to all ASAS-SN stars in our sample.
2. `reportable.csv`—The reported rotation periods for our best catalog of predictions with a fractional uncertainty below 25% along with ancillary Gaia DR2 and DR3 information. We provide a more detailed description of the information contained in this table in Table 4.
3. `training_clumpstars.csv`—The list of red clump stars we used to create our training set.
4. `allnetworkpredictions.zip`—The predicted rotation periods from all four neural networks explored in this work.
5. `asassn_cnn_model_weights.zip`—The model weights for all of the neural networks described in Section 4.

We also host all of the Jupyter Notebooks made in this work on Zenodo (M. Schochet et al. 2025b) at doi:10.5281/zenodo.17387593.

Additionally, all other code for this project is hosted on GitHub at https://github.com/m-schochet/asas-sn-cnn. We note that the entire ASAS-SN dataset—as either transformations or raw data—is substantially too large to host on any external platform; if these files are of interest, please reach out to the authors.

## Appendix A
## Injection Algorithm

Each of our ASAS-SN noise template light curves can be expressed as a list, Φ, which contains tuple values representing each data point.

$$\Phi = [\phi_i(\mathrm{t_i}),\ \phi_{ii}(\mathrm{t_{ii}}),\ ...,\phi_n(\mathrm{t_n})] \tag{A1}$$

where lowercase $\phi$ variables represent flux measurements, and t variables represent time stamps. Given that the `butterpy` simulations are created at 30 minutes cadence, the first step in our injection algorithm is to linearly interpolate the simulation fluxes to the time stamps of our noise light curve Φ. If we represent our `butterpy` simulations as Θ, where (times given in minutes)

$$\Theta = [\theta_1(t_1),\ \theta_2(t_2),\ ...,\theta_n(t_n)] \tag{A2}$$

$$t_n = t_1 + 30(n - 1);\quad n = 1, 2, 3... \tag{A3}$$

Then for each simulation–template pair, the full algorithm says that our interpolated flux, Γ, is

$$\Gamma = [\gamma_i(\mathrm{t_i}),\ \gamma_{ii}(\mathrm{t_{ii}}),\ ...,\gamma_n(\mathrm{t_n})] \tag{A4}$$

$$\gamma_n(\mathrm{t_n}) = \frac{\phi_n * [(\theta_n{}^*(\mathrm{t_{n+1}} - t_n)) + (\theta_{n+1}{}^*(t_n - \mathrm{t_n}))]}{\mathrm{median}(\Theta)^*(\mathrm{t_{n+1}} - t_n)}. \tag{A5}$$

## Appendix B
## "Chunky Lomb–Scargle" Algorithm

The following algorithm is taken essentially line by line from the `LS_wavelet` function found in the SkyPatrol GitHub (https://github.com/asas-sn/skypatrol/tree/master/pyasassn/wavelet.py) with minimal adjustments to variable names for readability.

```
import numpy as np
from astropy.timeseries import LombScargle

def LS_wavelet (times, frequencies, timestamps,
  fluxes, fluxerrors, tradeoff):
:param times: Array of times where the wavelet power
  should be evaluated
:param frequencies: Array of frequencies where the
  wavelet power should be evaluated
:param timestamps: Input time series time stamps
:param fluxes: Input time series dynamical measurement
  (fluxes in this scenario)
:param fluxerrors: Measurement uncertainties for the
  "fluxes" time series
:param tradeoff: Tradeoff between time and frequency
  resolution (has been preset to 2 throughout this work)

array = np.full(len(times), len(frequencies), np.nan)

def window(x):
return np.exp(-x**2/2)

for j, f in enumerate(frequencies):
dt = tradeoff * 1/f
for i, t in enumerate(times):
w = window((x-t)/dt)
m = np.isfinite(np.nan_to_num(fluxerrors/
  window, nan=np.inf))
ls = LombScargle(times[m], frequencies[m], dy=(flux-
  errors/window)[m])
p = float(ls.power(f, normalization="psd"))
p /= (np.sqrt(2 * np.pi) * dt)
array[i, j] = p

return array
```

## Appendix C
## Crossmatched Catalogs

For the analysis in Section 5.2, we present the number of overlapping crossmatched stars between archival works in Table 5.

**Table 5**
Overlap of Objects in Different Catalogs

| Catalog | Y. L. Lu et al. (2022) | A. R. G. Santos et al. (2020, 2021a) | Z. R. Claytor et al. (2024b) | Z. R. Claytor & J. Tayar (2025) | C. T. Christy et al. (2023) | R. J. Holcomb et al. (2022) | I. L. Colman et al. (2024) |
|---|---|---|---|---|---|---|---|
| Lu et al. | x | x | x | x | x | x | x |
| Santos et al. | ∼14 | x | x | x | x | x | x |
| Claytor et al. | ∼0 | ∼1 | x | x | x | x | x |
| Claytor and Tayar | 16 | 20,687 | ∼1 | x | x | x | x |
| Christy et al. | 369 | ∼61 | 89 | 49 | x | x | x |
| Holcomb et al. | ∼108 | ∼32 | 161 | ∼10 | 443 | x | x |
| Colman et al. | ∼43 | ∼34 | 216 | ∼11 | 496 | 4863 | x |

**Note.** Here we show overlapping stars between the catalogs that we compare to. Comparisons with integers are absolute overlaps, while comparisons with a "∼" sign indicate lower limits of overlapped objects (to account for any missed overlapping objects when matching by identifiers in different surveys).

## ORCID iDs

Meir E. Schochet https://orcid.org/0009-0007-9766-2324
Penelope Planet https://orcid.org/0009-0006-4199-3178
Zachary R. Claytor https://orcid.org/0000-0002-9879-3904
Jamie Tayar https://orcid.org/0000-0002-4818-7885
Adina D. Feinstein https://orcid.org/0000-0002-9464-8101